\documentclass[aps,pra,preprint,showpacs,groupedaddress,superscriptaddress,longbibliography]{revtex4-1}

\usepackage[english]{babel}
\usepackage{float}
\usepackage{graphicx}
\usepackage[caption=false]{subfig}
\usepackage{amsmath}
\usepackage{amsfonts}
\usepackage{amssymb}
\usepackage{mathtools}

\begin{document}

\title{Driven power-law oscillator}

 \author{Peter Schmelcher}
  \email{Peter.Schmelcher@physnet.uni-hamburg.de}
 \affiliation{Zentrum f\"ur Optische Quantentechnologien, Universit\"at Hamburg, Luruper Chaussee 149, 22761 Hamburg, Germany}
 \affiliation{The Hamburg Centre for Ultrafast Imaging, Universit\"at Hamburg, Luruper Chaussee 149, 22761 Hamburg, Germany}

\date{\today}

\begin{abstract}
We explore the nonlinear dynamics of a driven power law oscillator whose shape varies periodically
in time covering a broad spectrum of anharmonicities. Combining weak and strong confinement of
different geometry within a single driving period the phase space allows not only for regular 
and chaotic bounded motion but in particular also for an unbounded motion which exhibits 
an exponential net growth of the corresponding energies. Our computational study shows that phases of motion
with energy gain and loss as well as approximate energy conservation
alternate within a single period of the oscillator and can be assigned to the change of the underlying confinement geometry.
We demonstrate how the crossover from a single- to a two-component phase space takes place with varying frequency
and amplitude and analyze the corresponding volumes in phase space. In the high
frequency regime an effective potential is derived that combines the different features of the
driven power-law oscillator. Possible experimental realizations are discussed.
\end{abstract}

\maketitle

\section{Introduction} 
\label{sec:introduction}

Many physical systems can be modelled and described to some approximation by
individual oscillators or a collection of interacting
oscillators. Examples are vibrating molecules, phonons in solids and coupled Josephson junctions.
The route to complexity in the sense of a many-faceted dynamical behavior is then
at least two-fold: either one increases the number of (interacting) oscillators at the cost
of dealing with a high-dimensional generally mixed regular-chaotic phase space or one stays in low spatial dimensions and
accounts for a nonintegrable modification of the underlying integrable oscillator. The latter case has been
a paradigm not only for the route to chaos in low-dimensional systems but showcases many mechanisms 
of universal character that are valid also for higher-dimensional setups. In the present work
we will pursue this second route and introduce a novel type of one-dimensional driven oscillator
that unites properties usually occuring for different dynamical systems.
Before doing so and in order to provide a proper embedding and bottom-up approach in terms of
complexity let us briefly hint upon some important basic facts concerning oscillators and dynamical
billards in low dimensions.

For the one-dimensional harmonic oscillator exposed to dissipation and forcing closed form
analytical expressions are available \cite{Weinstock} for arbitrary time-dependent external forces.
In case of a periodic driving and for vanishing dissipation the motion is non-resonant, regular and bounded if the frequency
of the driving is unequal the one of the harmonic confinement. It is only for the resonant case of
equal frequencies that a linearly in time diverging amplitude is encountered. Nonlinear oscillators
such as the Duffing oscillator \cite{Kovacic,Korsch} or the kicked rotor (see \cite{Reichl,Stoeckmann} and references
therein) represent prototype systems whose phase space
decomposition varies from regular to mixed and finally predominantly chaotic with correspondingly
changing parameters. Specifically the kicked rotor exhibits a transition from a motion with bounded energy to
a diffusive dynamics. The latter leads to an unbounded momentum increase and a linear increase of the energy
in time takes place beyond a critical kicking strength. Switching from an external driving force
to a parametric driving the harmonic oscillator can gain energy for certain frequency ratios
of the natural frequency compared to the parametric driving frequency. The
maximum gain and exponential growth of the oscillation amplitude occurs for a ratio of one half and the 
oscillator then phase-locks to the parametric variation \cite{Fossen}. 

On the other hand side
Fermi acceleration (FA) \cite{Fermi} which refers to the unbounded growth of the energy of particles while repeatedly 
colliding with moving massive objects or fields has been of interest over many decades.
Prototype models showing FA are time-dependent hard wall two-dimensional billiards which have come into the particular focus 
of exponential FA during the past ten years \cite{Turaev1,Shah1,Liebchen,Shah2,Turaev2,Turaev3,Batistic}. 
Under certain conditions exponential FA can take place for most initial conditions. This has been 
demonstrated for the rectangular billiard with an oscillating bar \cite{Shah1,Liebchen,Shah2}
and for a class of chaotic billiards that exhibit a separation of ergodic components \cite{Turaev2,Turaev3}.
Importantly, it has been shown recently \cite{Batistic,Pereira} that under very general conditions
a generic time-dependent two-dimensional billiard exhibits FA in the adiabatic limit.
A key ingredient herefore is that the corresponding static, i.e. time-independent,
counterpart of the billiard exhibits a mixed regular-chaotic phase space.
However, the situation is very much different in a single spatial dimension. Here the prototype billiard
is the well-known Fermi-Ulam model describing a particle that multiply collides with moving walls \cite{Ulam}.
The Fermi-Ulam model does not allow for FA at all, and in particular not for the exponential FA, if the
applied time-dependent driving law is sufficiently smooth. This is due to the existence of invariant tori which
suppress the global energy transport \cite{Lieberman}. 

The current work goes one step further in the above line and introduces the
driven or time-dependent power-law oscillator (TPO).
Opposite to the above-mentioned cases, where the (harmonic) confinement 
of the oscillator remains intact during the external forcing, 
this is not any more the case for the TPO.
Here the confinement itself becomes time-dependent or, put it another way,
the exponent of the power-law potential becomes time-dependent. Specifically we focus on the
case of a periodic driving characterized by a frequency and an amplitude as well as a constant offset.
Within a single driving period phases of strong and weak anharmonic confinement follow upon each other thereby
covering a continuous interval of power law confinement strengths.
Our aim is to provide a first computational study of this driven power-law oscillator with an emphasis
on the analysis of its phase space. The latter is straightforwardly possible due the low dimensionality
of the TPO. As we shall see this peculiar oscillator provides an unusual combination of dynamical
features which one might not suspect to appear from a first glance at its
Hamiltonian and from the preexperience gained from the 'traditional' oscillators.
Obviously the TPO characterized by the time-dependent 'pumping' of its anharmonicity is very much different from the billiard
dynamics. Nevertheless the question for a comparison of the
their energy growth behavior is an intriguing one.

In detail we proceed as follows. In section \ref{sec:setup} we introduce our setup and in particular
the underlying time-dependent model Hamiltonian for the TPO. We discuss the properties of
the instantaneous one-dimensional potential whose confinement behavior changes qualitatively for different
times. The underlying phase space is equally considered. Section \ref{sec:results} provides an analysis of the dynamics
first by inspecting trajectories and second by investigating the Poincare surfaces of section. This allows us to gain
an overview of the dynamical components in phase space. A quantification
of the volume of the different components of the phase space is provided. Subsequently the time evolution
of the energy behavior for an ensemble is provided.
This section contains also a discussion of the high frequency regime. We present our conclusions
and outlook in section \ref{sec:conclusions}.

\section{Setup, Hamiltonian and Static Phase Space} \label{sec:setup}

The TPO as defined below is motivated by the idea that a most general oscillator would
also allow for a time-dependent variation of the geometry of the spatially confining potential. This is obviously not
the case for the harmonic oscillator exposed to an external forcing, the Duffing oscillator or the
parametric oscillator with a time-dependent frequency. The most straightforward possibility to change the
shape of the potential time-dependently is to maintain its power-law appearance but make the exponent
time-dependent. The TPO Hamiltonian then reads

\begin{equation}
{\cal{H}} = \frac{p^2}{2} + \alpha q^{2 \beta(t)} \label{eq1}
\end{equation}

and contains the time-dependent potential $V(q,t)= \alpha q^{2 \beta(t)} := \alpha ({q^{2}})^{\beta(t)}$. The latter
is a defining equation which ensures a positive real argument and
implies that always the positive real root is taken for arbitrary powers $\beta$.
We put for reasons of simplicity the mass of the oscillator to one.
Our focus in this work is on a periodic time-dependence of the exponent $\beta(t)=\beta_0+
\beta_1 sin(\omega t)$ which contains the amplitude $\beta_1$ of the sinusoidal driving and the constant offset $\beta_0$.
$\alpha =1, \beta_0 = 1$ will be chosen throughout which means that the oscillation takes place around
the harmonic oscillator potential $\alpha q^2$ as an equilibrium configuration.
Depending on the amplitude $\beta_1$ the TPO possesses alternating phases of strong and weak anharmonic confinement
covering continuously the noninteger powers within an individual oscillation.

\begin{figure}
\includegraphics[width=6cm,height=6cm]{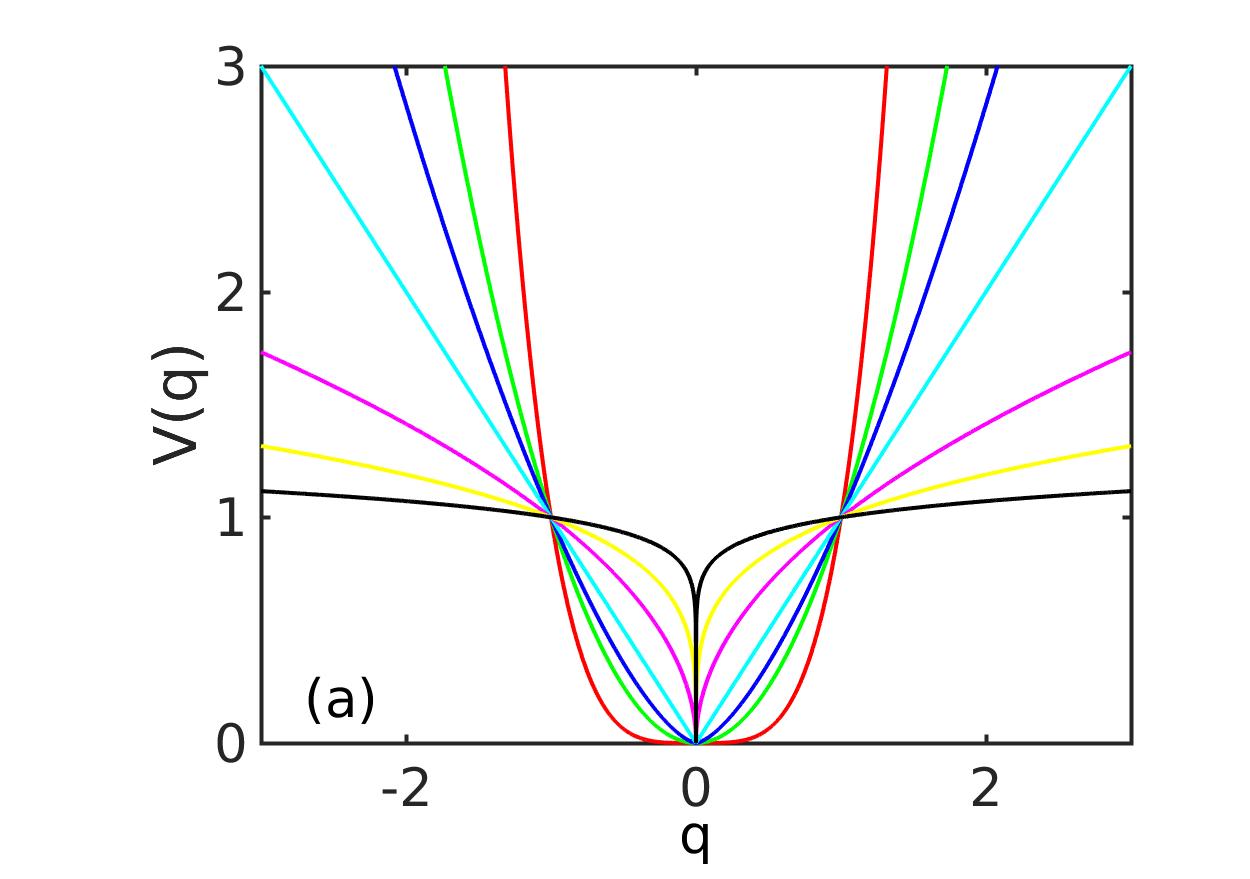} 
\includegraphics[width=6cm,height=6cm]{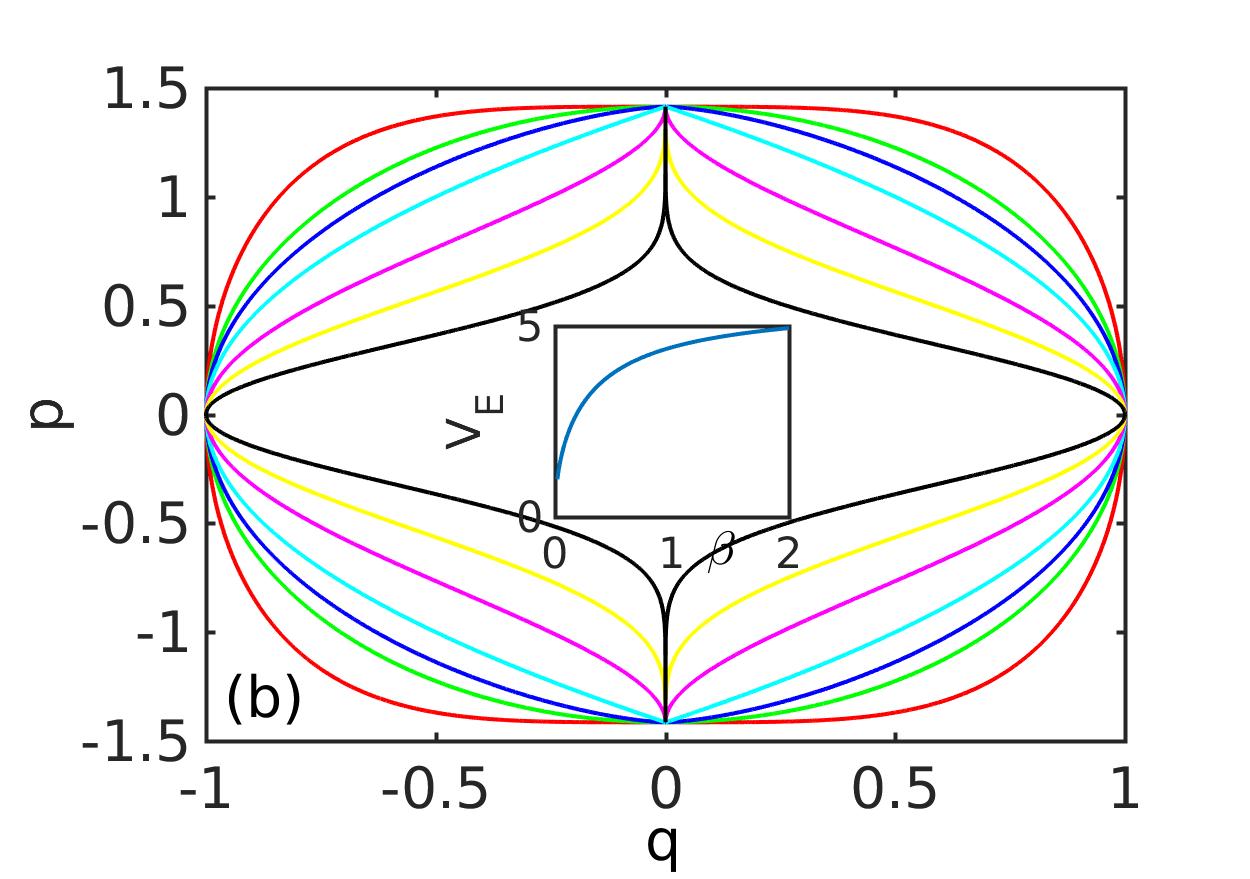}
\caption{\label{fig:fig1} (a) Snapshots of the potential $V(q)$ for different instantaneous values of the parameter
$\beta$. Red, green, blue, cyan, magenta, yellow and black lines correspond to the values $\beta=2,1,0.75,0.5,0.25,0.125,0.05$
ordered according to their appearance from top to bottom for $q>1$. (b) Phase space (q,p) curves
for the same values of $\beta$ following the identical colour coding.
The energy chosen is $E_0=1$. The central inset shows the volume $V_E$
of the phase space for energies $E \le E_0$ as a function of the parameter $\beta$ (see text).}
\end{figure}

It should be noted that, due to the occurence of arbitrary fractional powers already in the Hamiltonian (\ref{eq1})
an analytical approach of whatever kind is in general not obvious. The reason herefore is that the
resulting integrals cannot be evaluated analytically.
For the numerical solution of the resulting equations of motion we employ a
fourth order Adams-Moulton predictor corrector integrator. However, one has to take into account
the singularties of the potential and its derivates. Indeed $V(q,t)$ becomes singular
at $q=0$ for $\beta(t) <0$ if $\beta_1 > \beta_0$. Equally the first and second derivative
become singular at $q=0$ for $\beta_1 > \beta_0 - \frac{1}{2}$ and $\beta_1 > \beta_0 - 1$, respectively.
Therefore a regularization of the equations
of motion is appropriate. We accomplish this by replacing the potential $V(q,t)= \alpha q^{2 \beta(t)}$ by its regularized
version $V(q,t) = \alpha \left(q^2 + C \right)^{\beta(t)}$. $C$ is chosen to be 
a very small (typical value of the order of $10^{-8}$) positive constant removing all singularities at
the origin. The impact of this tiny regularization constant on the actual motion is very limited 
and controllable. For the case $\beta \gtrsim 1$ one can estimate the impact of $C \ne 0$ on the dynamics
by calculating the relative change of the velocity in a corresponding 'collision'. This amounts
to $\frac{\Delta v}{v} = 2 \alpha \beta \frac{C^{\beta}}{v^2}$ which is for small $C$ and a very large 
range of velocities a tiny correction. Moreover, for $\beta << 0.5$ one can show that the spatial
extension $\Delta x = 2 \cdot \left( (\frac{y}{\alpha})^{\frac{1}{\beta}} - C\right)^{\frac{1}{2}}$ on which
the potential acquires values $y<<0.1$ decreases rapidly to zero,
which renders the impact of a correspondingly small $C \ne 0$ also 
very small (see Figure \ref{fig:fig1}(a) for the cusp at the origin).

Since the TPO (\ref{eq1}) is of quite unusual appearance let us first discuss the instantaneous 
form of the potential as the time-evolution proceeds. This evolution is illustrated in Figure \ref{fig:fig1}(a).
For values $\beta >1$ the nonlinear confinement is stronger (for $q>1$) than that of the harmonic oscillator covering
in time a continuous range of exponents $1 < \beta < 1+\beta_1$ where the curvature is always positive.

For values $0.5 < \beta <1$, however, the curvature decreases and for $\beta=0.5$ $V(q)$ becomes linear.
Further decreasing its value ($0 < \beta < 0.5$) the potential shows now a very weak confining behavior
and exhibits a negative curvature. Here the formation of a negative cusp
of $V(q)$ can be observed (note that $V(0)=0$ for $C=0$ holds also in this regime). Due to this
cusp the dynamics close to the origin of such a (static) potential experiences kicks i.e. sudden
changes of the underlying momentum. How much of the above
qualitatively very different behavior of the instantaneous potential $V(q)$ is covered for a 
corresponding TPO (\ref{eq1}) depends of course on the value of the amplitude $\beta_1$. 

The corresponding instantaneous phase space is illustrated in Fig.\ref{fig:fig1}(b). For $\beta>1$ the 
phase space curves are convex but for $\beta<1$ they develop a cusp at $q=0$ and become piecewise concave.
The total phase space volume bounded by the energy shell with energy $E_0$ i.e. the volume of the
set $\{(q,p)| E(q,p) \leq E_0 \}$ for a fixed $E_0$, is given by $V_E = \int_{{\cal{H}}(q,p) \le E_0} dp dq$.
Note that this quantity represents in ergodic systems an adiabatic invariant \cite{Robnik}.
$V_E$ is shown in the central inset of Fig.\ref{fig:fig1}(b) and we
observe that it shrinks monotonically with decreasing $\beta$.
In particular also the average width in momentum space decreases with decreasing
$\beta$. Close to maximal momenta are then only accessible in the vertical channels around $q=0$ which
become increasingly narrow with decreasing $\beta$. 

\section{Results and Discussion} \label{sec:results}

Due to the three-dimensional (q,p,t) character of our system a complete overview of phase space can be achieved
by performing stroboscopic snapshots in time for each period $T=\frac{2 \pi}{\omega}$. One
might expect that the intermediate frequency regime (for which the driving frequency $\omega$ and the natural
frequency of the time-independent oscillator are comparable) is the most interesting. Then a mixed
phase space of regular and chaotic motion, depending on the size of the oscillation amplitude, could be
expected. The high frequency 
and low frequency motion could be predominantly regular at least for small 
amplitude. A natural question to pose is: how does the energy transfer to the oscillator take place and
what is the time-evolution of its energy. 
Does the energy stay bounded or become unbounded, and if the latter occurs, is the increase of
power-law or of exponential character ? From the quasistatic picture discussed in section \ref{sec:setup}
one might conjecture that during the half-period of each cycle of the driving for which $\beta > \beta_0$ the 
strong confinement prevents the oscillator from attaining a large amplitude and therefore energy and
vice versa for the other half-cycle for which $\beta < \beta_0$. 

\subsection{A first glance at the dynamics: Trajectories} \label{sec:trajectories}

In order to gain a first insight into the dynamical behavior of the driven power-law
oscillator it is instructive to consider individual trajectories. The decisive
parameters to be varied for fixed $\beta_0 =1, \alpha = 1$ are the amplitude of the oscillation $\beta_1$
and the underlying frequency $\omega$.

\begin{figure}
\includegraphics[width=5cm,height=6cm]{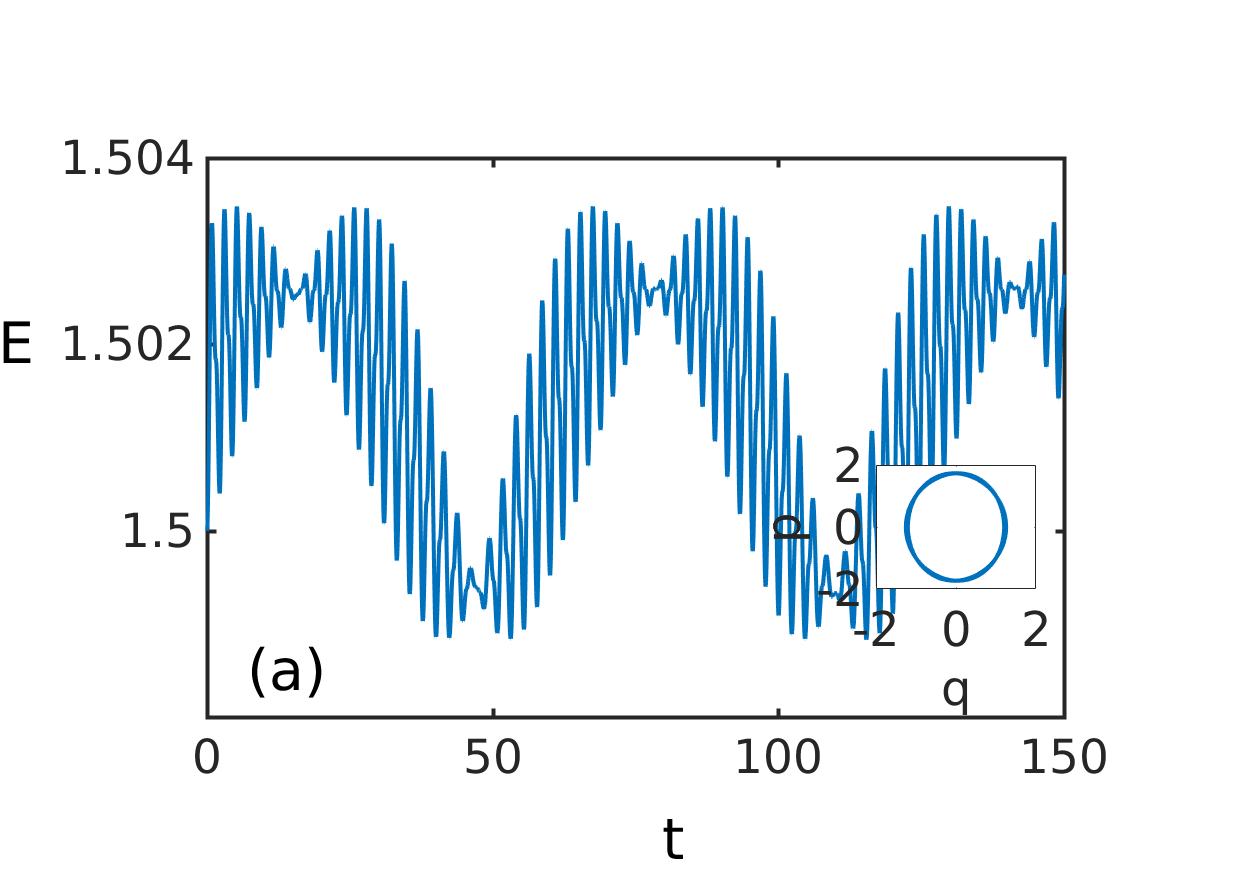} 
\includegraphics[width=5cm,height=5.5cm]{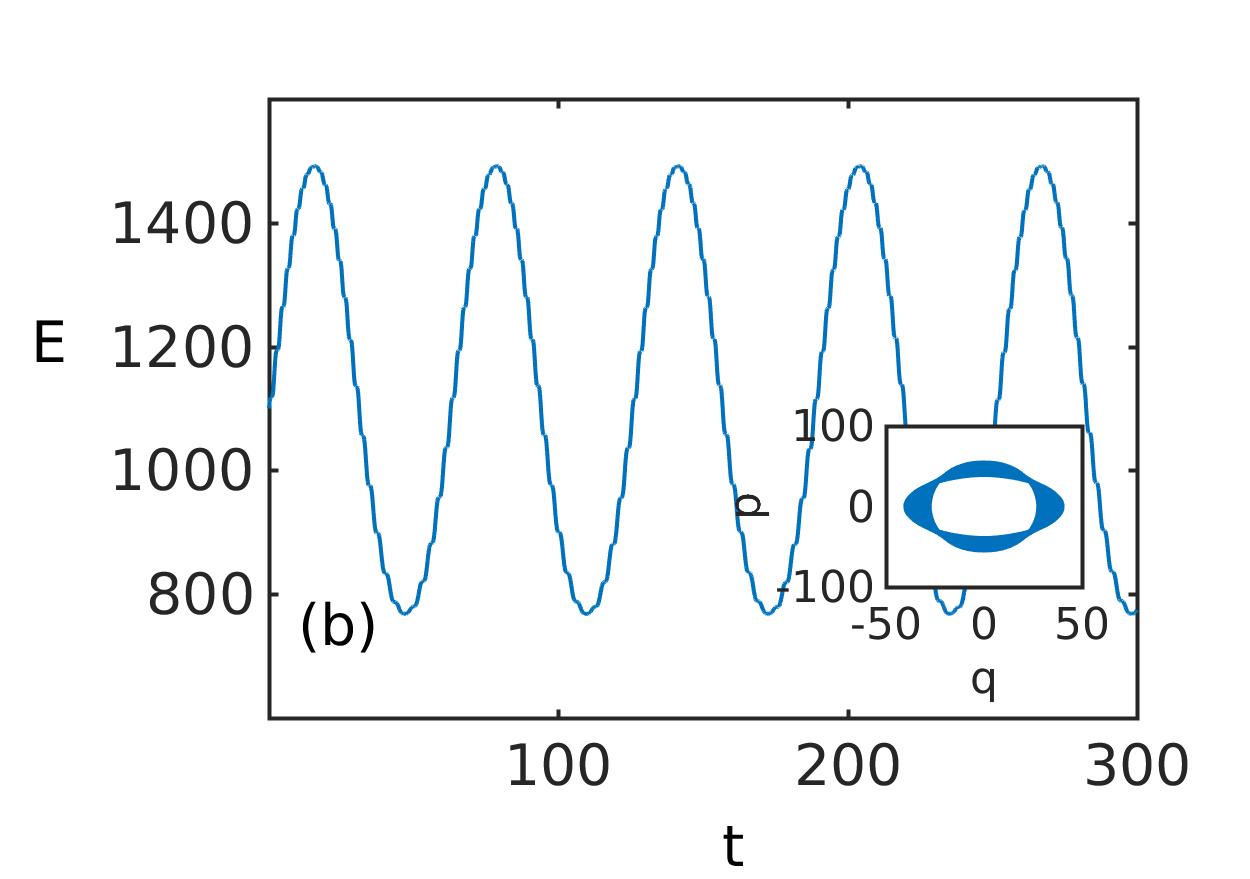}
\includegraphics[width=5cm,height=5.3cm]{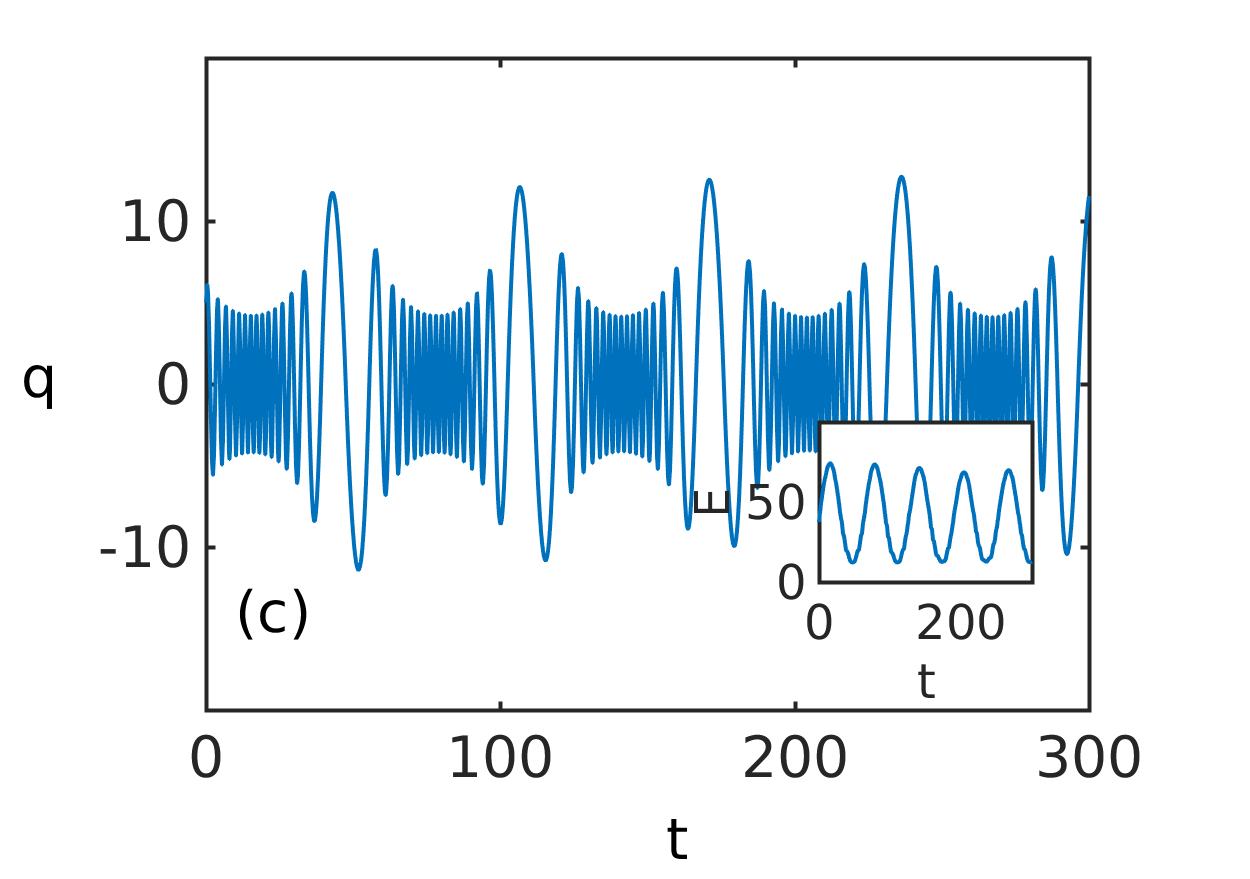}
\includegraphics[width=5cm,height=5cm]{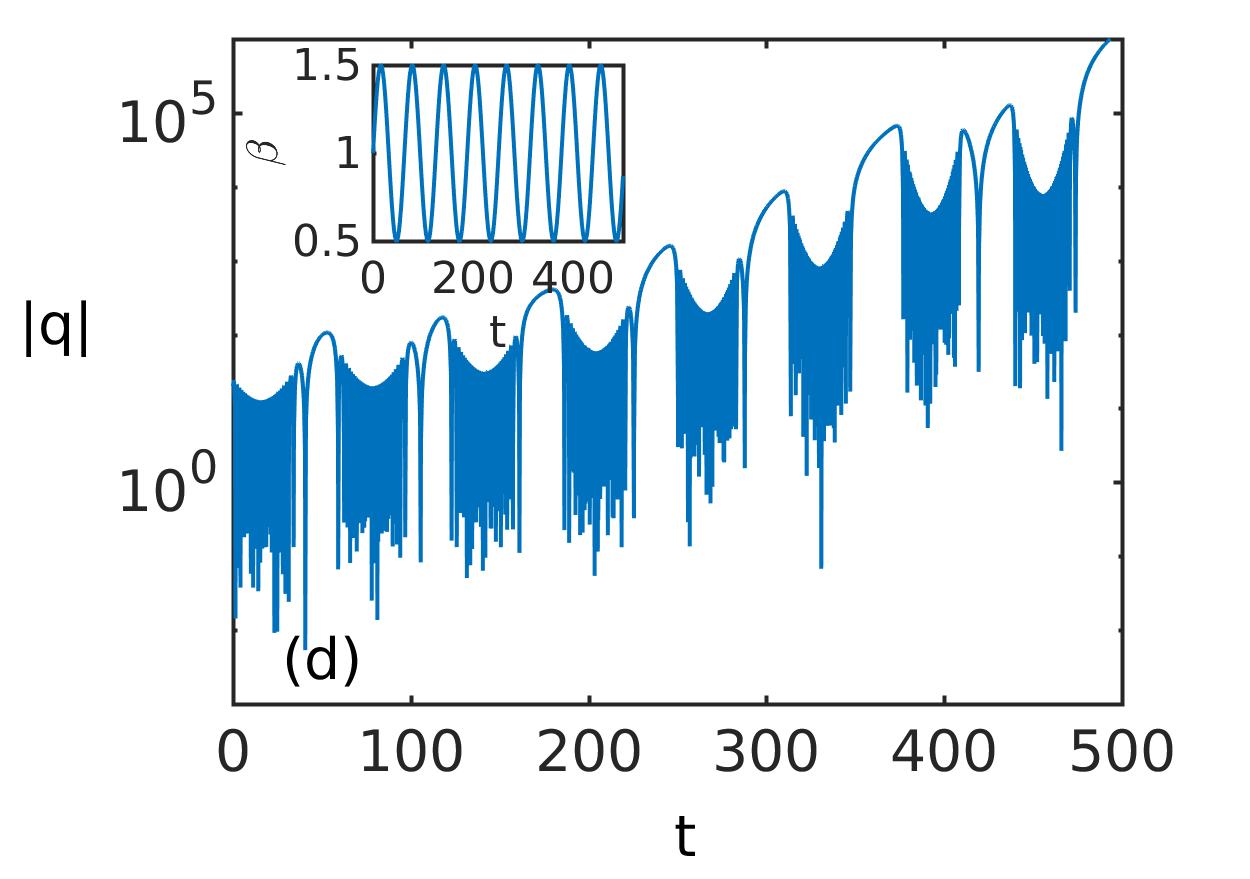} 
\includegraphics[width=5.4cm,height=5.5cm]{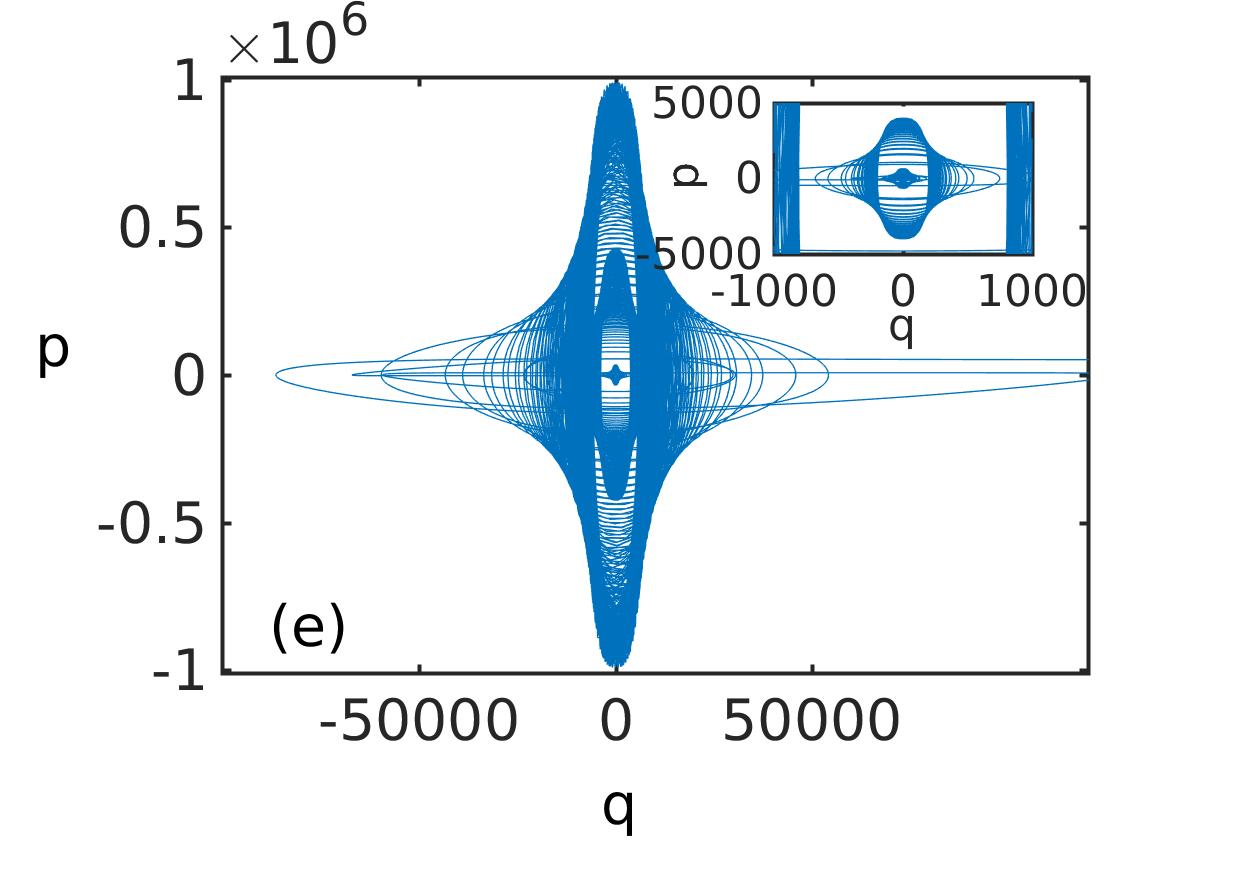}
\includegraphics[width=5cm,height=5cm]{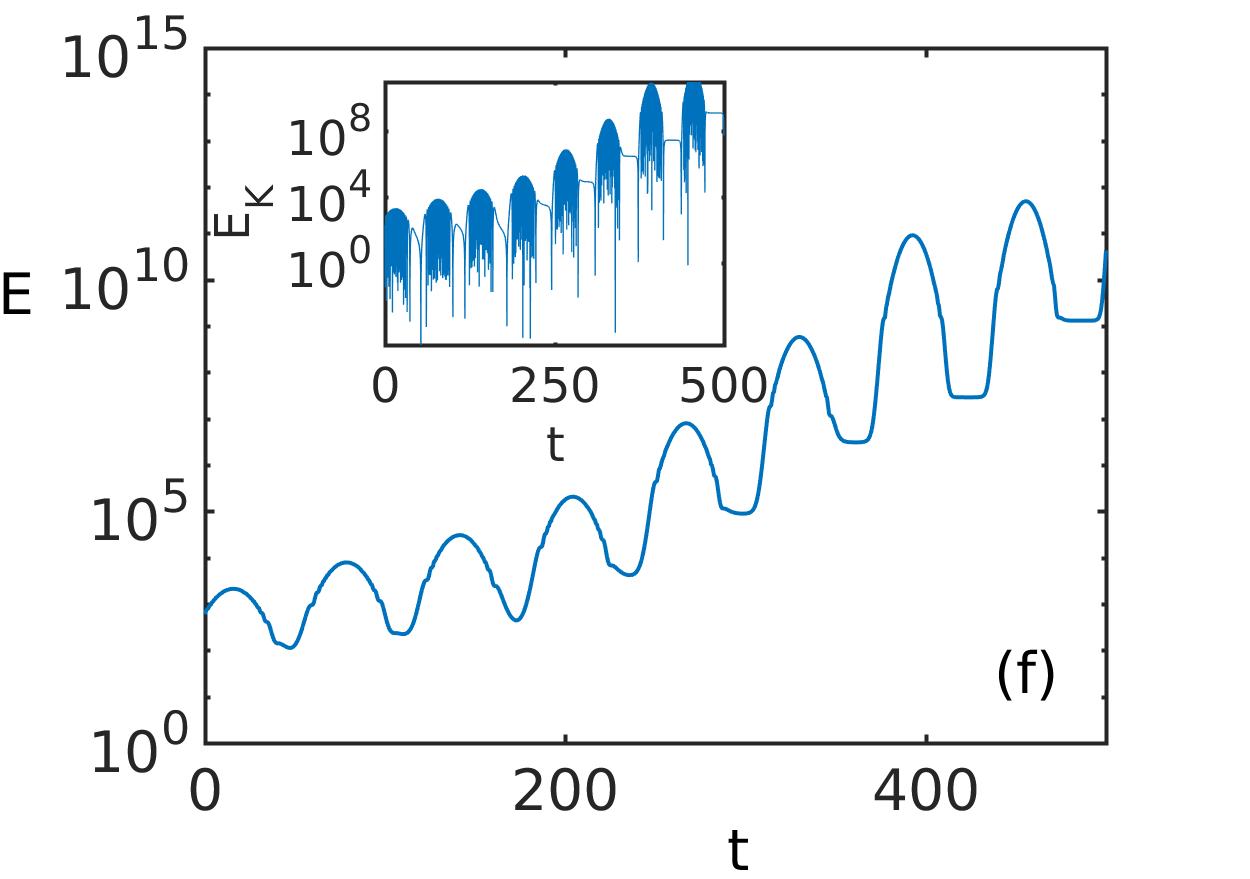}
\caption{\label{fig:fig2} Trajectories of the TPO for $\omega = 0.1$ with
varying amplitude $\beta_1$. (a) shows the total energy $E(t)$ as a function of time and in the inset the covered phase
space $(q(t),p(t))$ during the same time period for the IC $q(t=0)=1,p(t=0)=1$ for $\beta_1=0.1$. (b) Same as for
(a) but for the IC $q(t=0)=-30,p(t=0)=20$. (c) Amplitude $\beta_1=0.5$; the time-evolution of the coordinate $q(t)$ 
and in the inset the corresponding total energy $E(t)$ for the IC $q(t=0)=5,p(t=0)=5$. 
(d) $\beta_1 = 0.5$: the time evolution of the absolute value $|q|(t)$ for the
IC $q(t=0)=20,p(t=0)=20$ together with $\beta(t)$ in the inset. (e) same parameter values as in (c,d), but
showing the corresponding phase space curves $(q(t),p(t))$ on different scales (zoom in for inset)
and (f) showing the corresponding time evolution of the total (main figure) and kinetic energies (inset) for the same IC.}
\end{figure}

Let us briefly discuss some relevant aspect concerning the initial conditions (IC) of the trajectories for given
parameters of the oscillator. For the time-independent oscillator for arbitrary values of $\beta$
the distance of the IC from the origin $(q=0,p=0)$ in phase space plays no essential
role for the resulting phase space curves. Their overall shape remains the same.
However, as we shall see below, this distance is crucial for the individual trajectory dynamics of
our TPO. Therefore, we shall choose in the following as examples ICs close and far from the origin.
The origin itself is a fixed point of the dynamics in all cases of the TPO, i.e. for
arbitrary parameter values, while it might change its stability character (see below). The latter is not possible for
the time-independent oscillator where this fixed point is always stable for $\beta > 0$. 

Let us begin with the case $\beta_1=0.1, \omega=0.1$ which is the case of comparatively small amplitude
and low frequency. Figure \ref{fig:fig2}(a) shows for the IC $(q(t=0)=1,p(t=0)=1)$, i.e. close to the origin, the time-dependence of the energy.
It exhibits two frequencies. The smaller one corresponds to the driving frequency
$\omega$ and the larger one to the instantaneous oscillator frequency. The latter varies not significantly due to
the small amplitude $\beta_1$ in this case. Note that the overall variation of the energy is very small
and amounts to less than a percent of the initial energy $E(t=0)$. The inset in Figure \ref{fig:fig2}(a)
shows the corresponding phase space curves which, on the given time scale, shows small deviations from
the time-independent case. In Figure \ref{fig:fig2}(b) for the same setup the IC $(-30,20)$ is chosen,
i.e. an IC far from the central fixed point. Here we observe a large variation of the energy with time
of more than $50 \%$ of the initial energy. Still the energy shows bounded oscillations with essentially
two frequencies. The smaller one again corresponds to the driving frequency $\omega$ and the larger one (note that the
corresponding oscillation is hardly visible in Figure \ref{fig:fig2}(b) due to the small amplitude) 
to the instantaneous oscillator or confinement frequency. The corresponding phase space curves in
the inset show now significant deviations from the initial winding
of the phase space trajectory thereby changing its overall anisotropic shape.

The dynamics changes qualitatively if we consider the situation of an increased amplitude. Figures \ref{fig:fig2}(c-f)
analyze the dynamics again for a low frequency $\omega = 0.1$ but a substantially increased amplitude
$\beta_1=0.5$ such that the oscillator potential changes between $|q|$ and $|q|^{3}$ (assuming $C=0$) which covers
a broad interval of anharmonicities of fractional powers. Figure \ref{fig:fig2} (c) shows the time
evolution in coordinate space $q(t)$ for the IC $(5,5)$. The envelope behavior obeys the frequency $\omega=0.1$ but  
now a strong chirp of the frequency of the individual oscillations can be observed. Large amplitude behavior
occurs for low frequencies whereas small amplitudes occur at high frequencies. The inset demonstrates
that a large variation of the underlying energy $E(t)$ takes place at the frequency $\omega$ (the larger
frequency and small amplitude oscillations are hardly visible in the inset). We emphasize that all the
dynamics observed so far is regular, i.e. periodic and quasiperiodic, and in particular bounded which holds
also for the corresponding energy.

The above situation changes if we consider an IC $(20,20)$ further
away from the stable elliptic fixed point of the TPO at the origin. Figure \ref{fig:fig2} (d) shows $|q|(t)|$ on a 
semilogarithmic scale as a function of time for a somewhat longer time interval as compared to 
Figure \ref{fig:fig2} (c). Here we observe in the average an exponential increase of the amplitude of
the coordinate oscillations which equally cover positive and negative values. Periods of high frequency
and smaller amplitude motion are intermittently interrupted by periods of lower frequency large amplitude motions.
With the help of the corresponding inset showing $\beta(t)$ the latter can be assigned to the overall
perodic oscillations of the TPO. For time intervals where the power of the oscillator covers the interval
$1 \le 2 \beta \le 2$, i.e. if the oscillator confinement weakens from quadratic to linear, large amplitude
excursions occur. For time intervals with $2 \le 2 \beta \le 3$, i.e. if the oscillator confinement increases
up to a cubic behavior, the high frequency small amplitude oscillations are encountered.
Figure \ref{fig:fig2} (e) shows the corresponding phase space curves $(q(t),p(t))$. Extending on the 
discussion of the anisotropically evolving phase space curves, according to the inset of
Figure \ref{fig:fig2}(b), we now observe that this pattern repeats subsequently on different length scales.
This is shown in Figure \ref{fig:fig2}(e) and its inset for a zoom into the central part of the phase space. According to the 
observed exponential scaling the winding of the phase space curves at high frequency on their individual length scales
are connected by low frequency intermediate transients which connect the different length scales. The dynamics therefore
repeats as time proceeds on different length scales in a self-similar fashion, which can nicely be seen in 
Figure \ref{fig:fig2}(e). Finally Figure \ref{fig:fig2}(f) shows the time evolution of the corresponding
total energy $E(t)$ and kinetic energy $E_K(t)$ in the inset on a semilogarithmic scale.
During the time intervals of low frequency high
amplitude motion of $|q|(t)$ (see Figure \ref{fig:fig2}(d)) a plateau-like behavior is observed for the
energy $E(t)$. During the high frequency small amplitude motion of $|q|(t)$ a strong increase and decrease of the total
energy can be observed. $E(t)$ increases strongly when the amplitude of the high frequency oscillations decreases
and vice versa. In the average the total energy increases exponentially.
As a matter of precaution, we note that this and similar statements are to be understood
within our finite time simulations of a finite portion of the TPO phase space.
A similar behavior can be observed for the kinetic energy $E_K$ in the inset
of Figure \ref{fig:fig2}(f) which demonstrates in particular that also the kinetic energy grows in the
average exponentially. The kinetic energy becomes large during intervals of motion of the TPO with strong
confinement $2 \le \beta \le 3$ i.e. when the TPO compresses and 'pumps' the motion. Low kinetic
energies occur for the intervals of motion $1 \le \beta \le 2$ of the TPO where it releases the motion
to a much less confining potential. The latter interval coincides with the plateau-like behavior of the
total energy, i.e. we have approximately no increase or decrease of the total energy.

Already the above-analyzed few trajectories for given parameter values 
exhibit a rich behavior of the TPO dynamics. Having this in mind, let us now gain an overview of complete 
phase space by inspecting the corresponding Poincare surfaces of section (PSS). These PSS are stroboscopic snapshots
of the dynamics at periods of the external driving. We will explore the PSS with varying parameters.

\subsection{Poincare surfaces of section} \label{sec:psos}

Our above-analyzed few example trajectories have been focusing on the low frequency case. We have 
observed that with increasing amplitude $\beta_1$ of the TPO there exist not only regular trajectories with 
bounded energy oscillations, but also trajectories that exhibit an exponential increase in position and
energy. In the following we discuss the corresponding PSS of our TPO for different frequencies $\omega$ and
amplitudes $\beta_1$. This way we will gain a more complete overview of phase space with varying parameters. Note, that
due to the fact that the phase space of the TPO is unbounded from above, the following computational studies focus
on a finite but quite large energy window above zero energy. The minimal zero energy is given by the fixed point at the origin
(assuming $C=0$).

\begin{figure}[H]
\includegraphics[width=5cm,height=5cm]{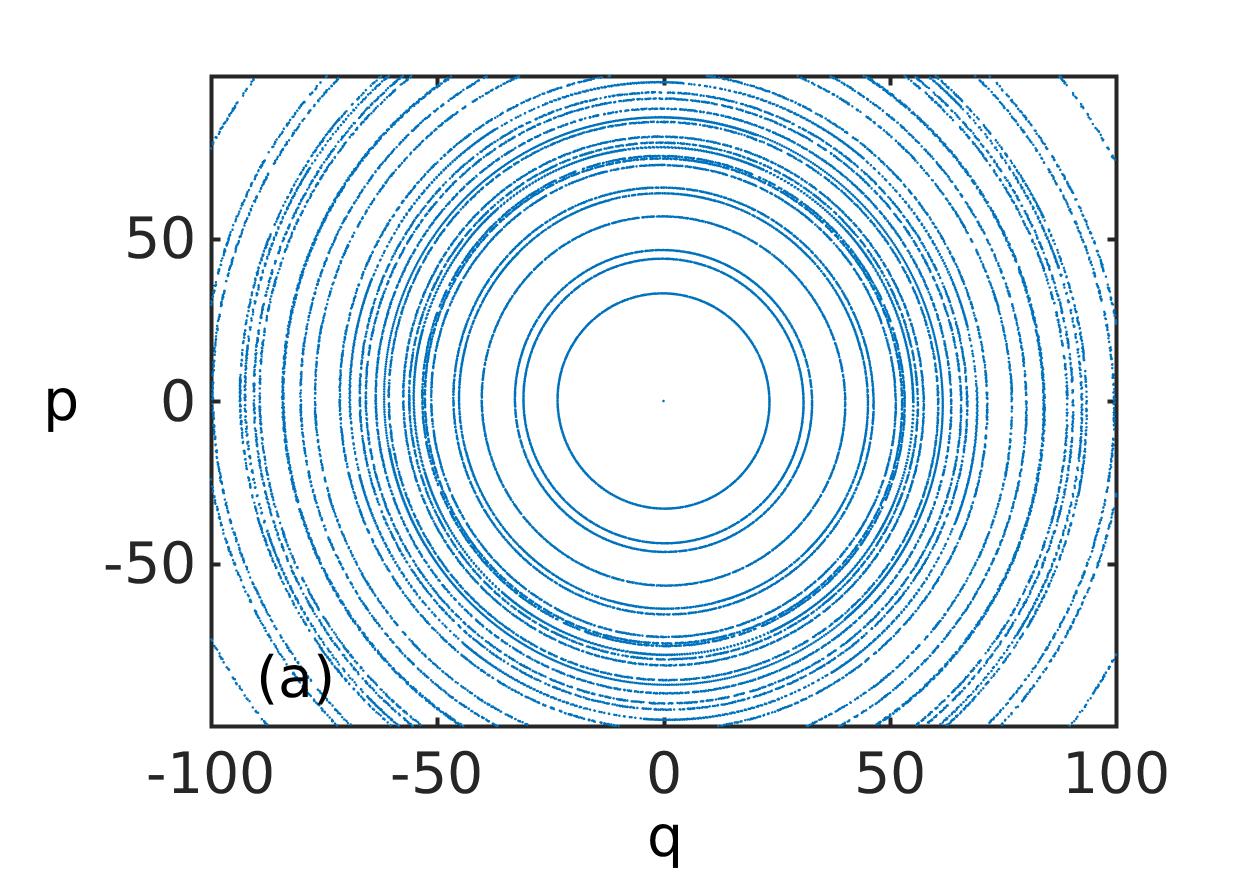} 
\includegraphics[width=5cm,height=5cm]{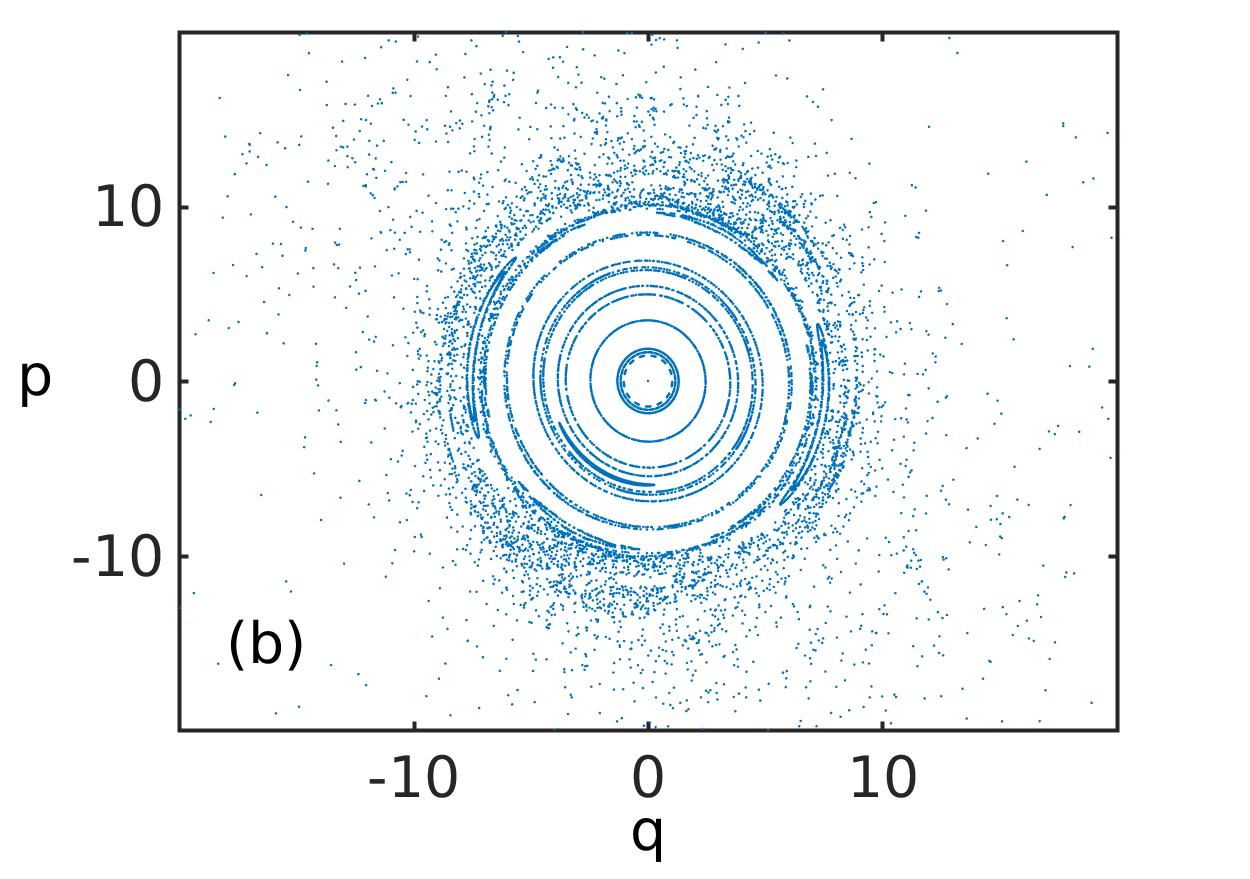}
\includegraphics[width=5cm,height=5cm]{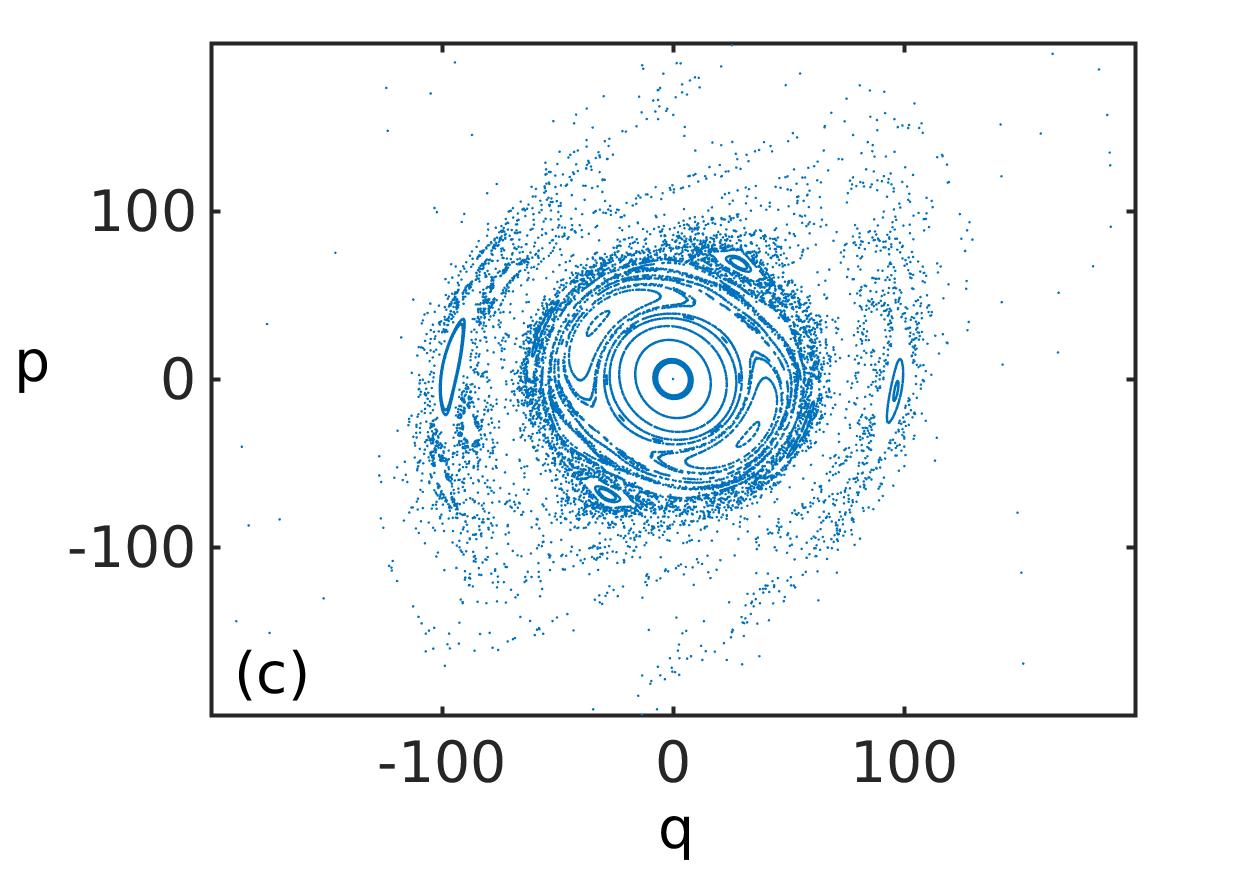}
\includegraphics[width=5cm,height=5cm]{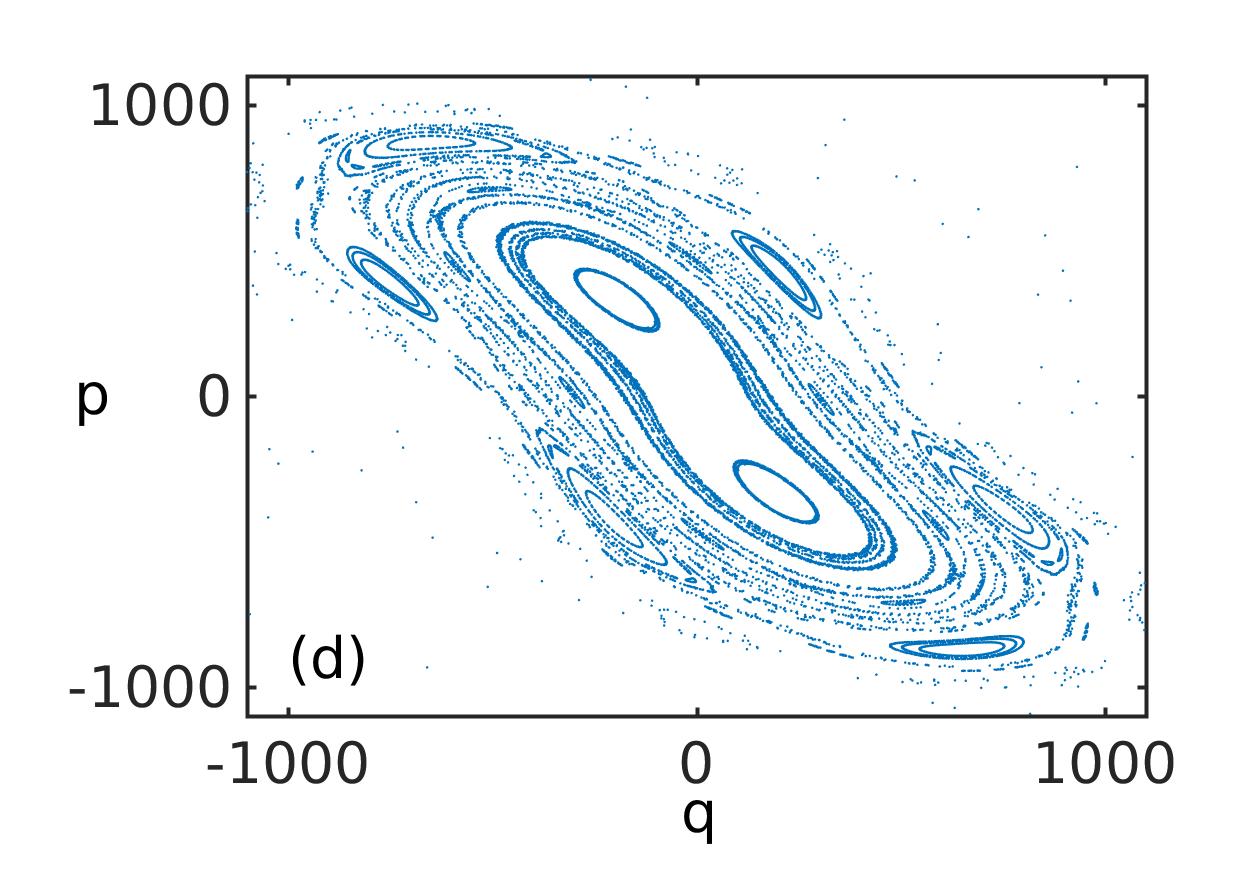}
\includegraphics[width=5cm,height=5cm]{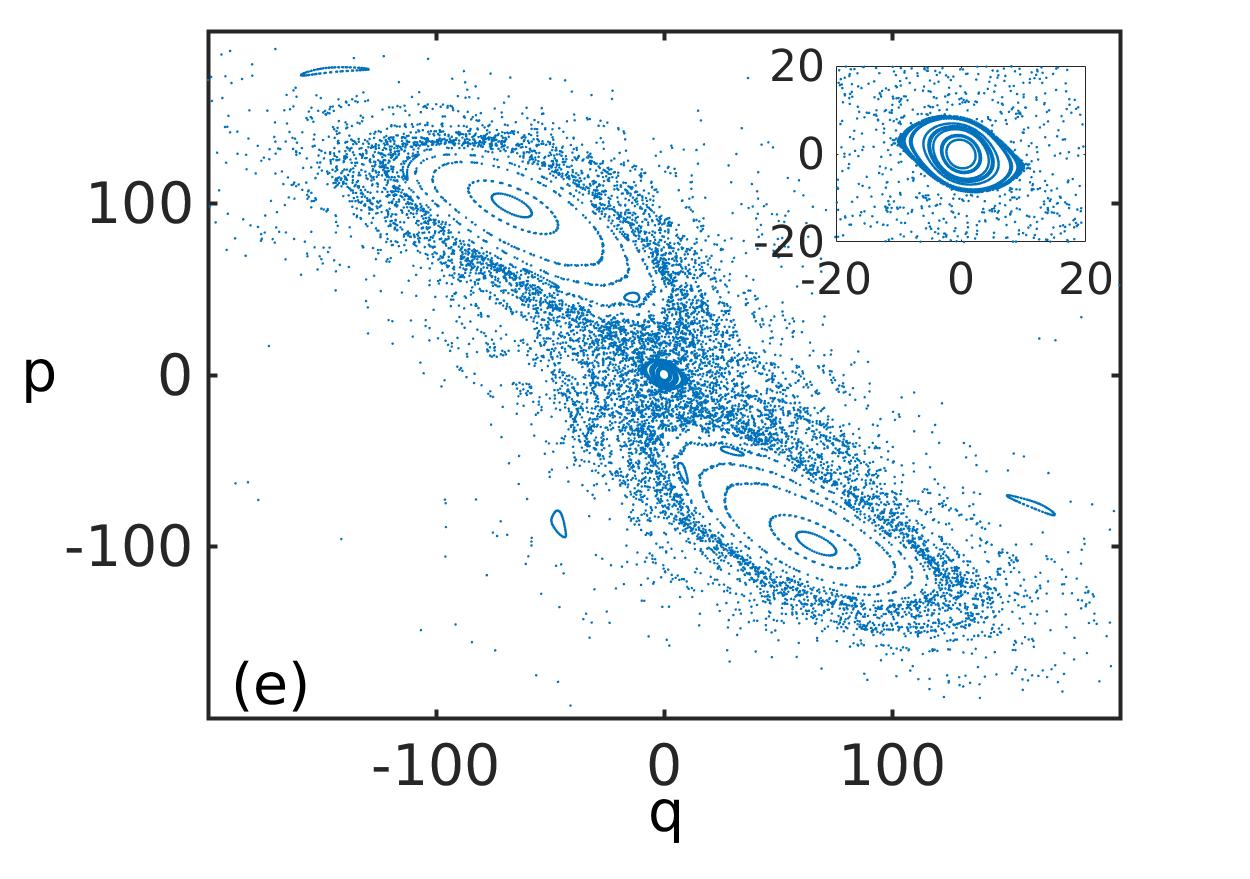}
\includegraphics[width=5cm,height=5cm]{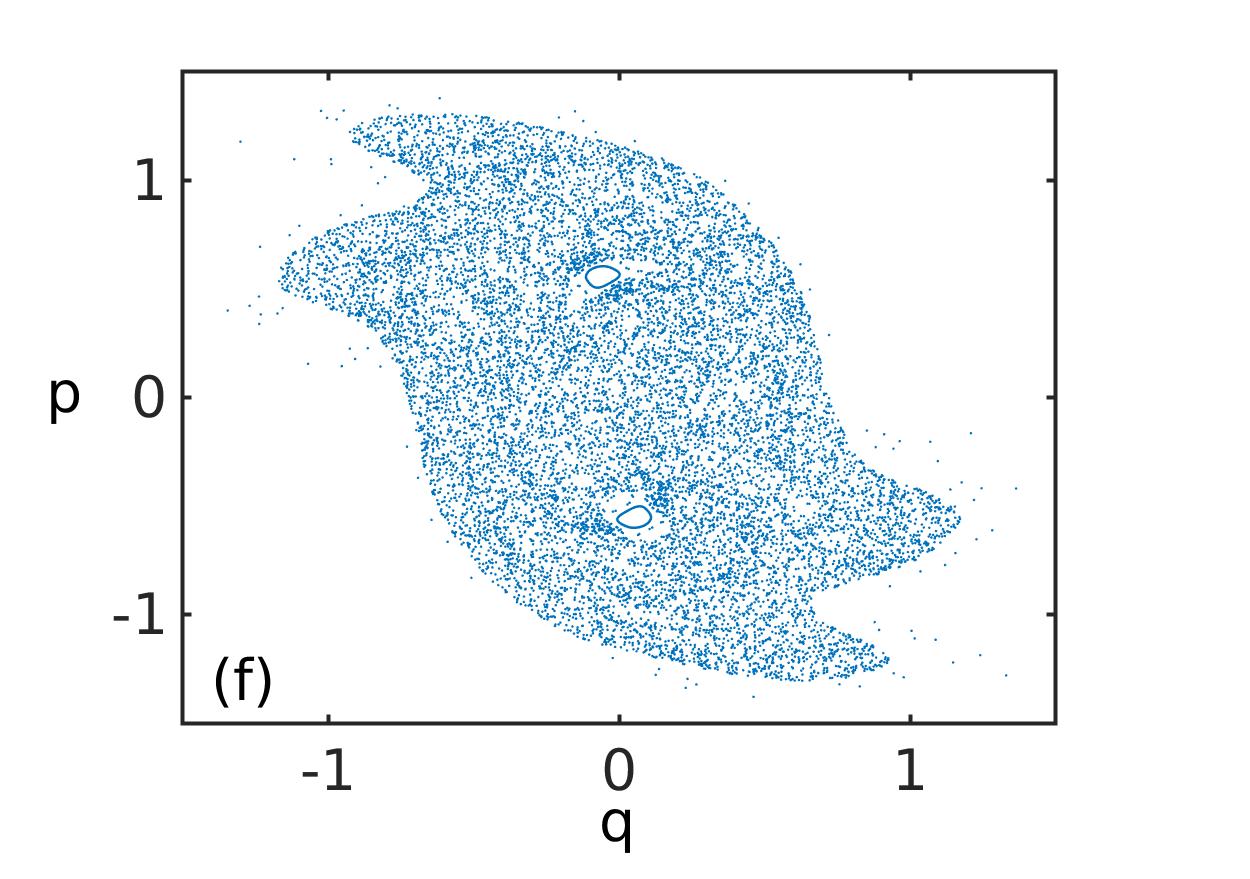}
\includegraphics[width=5cm,height=5cm]{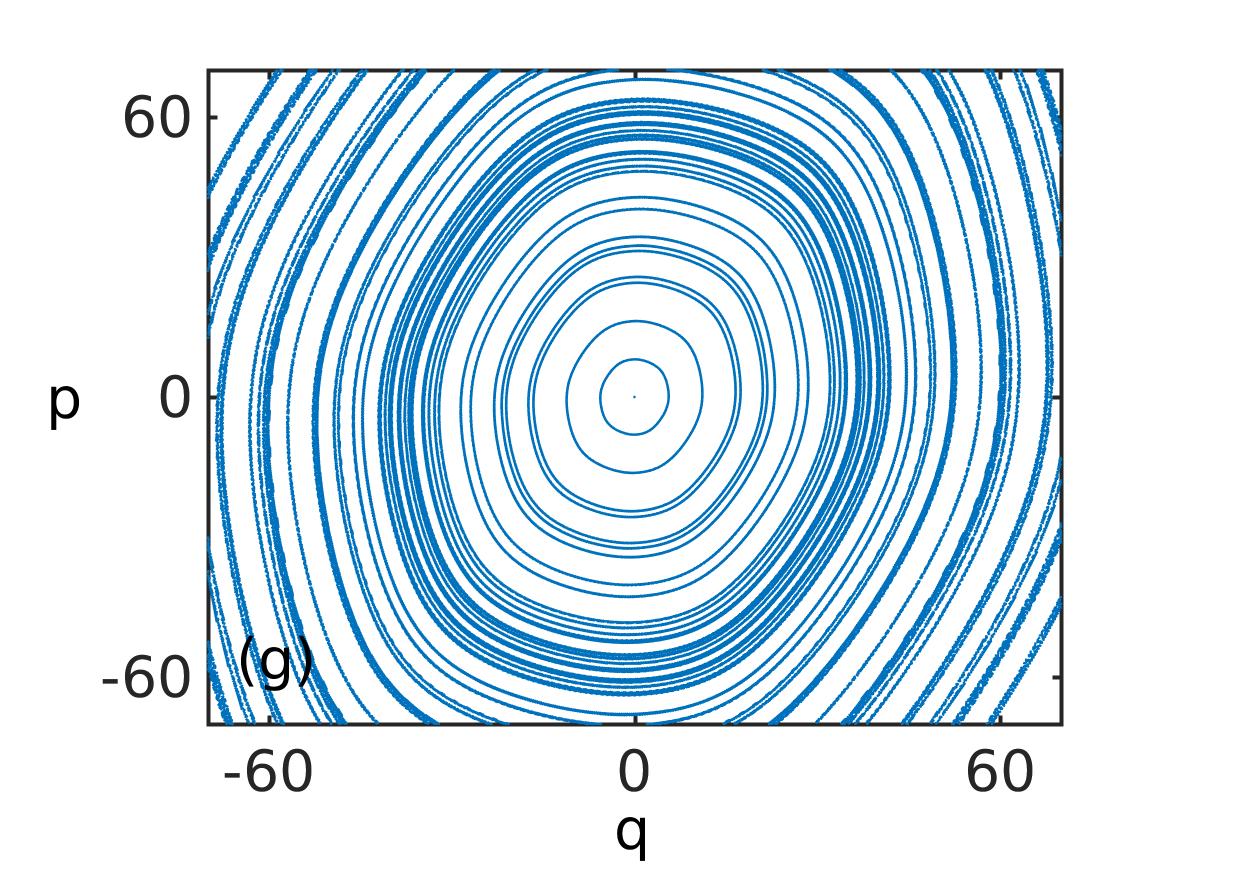}
\includegraphics[width=5cm,height=5cm]{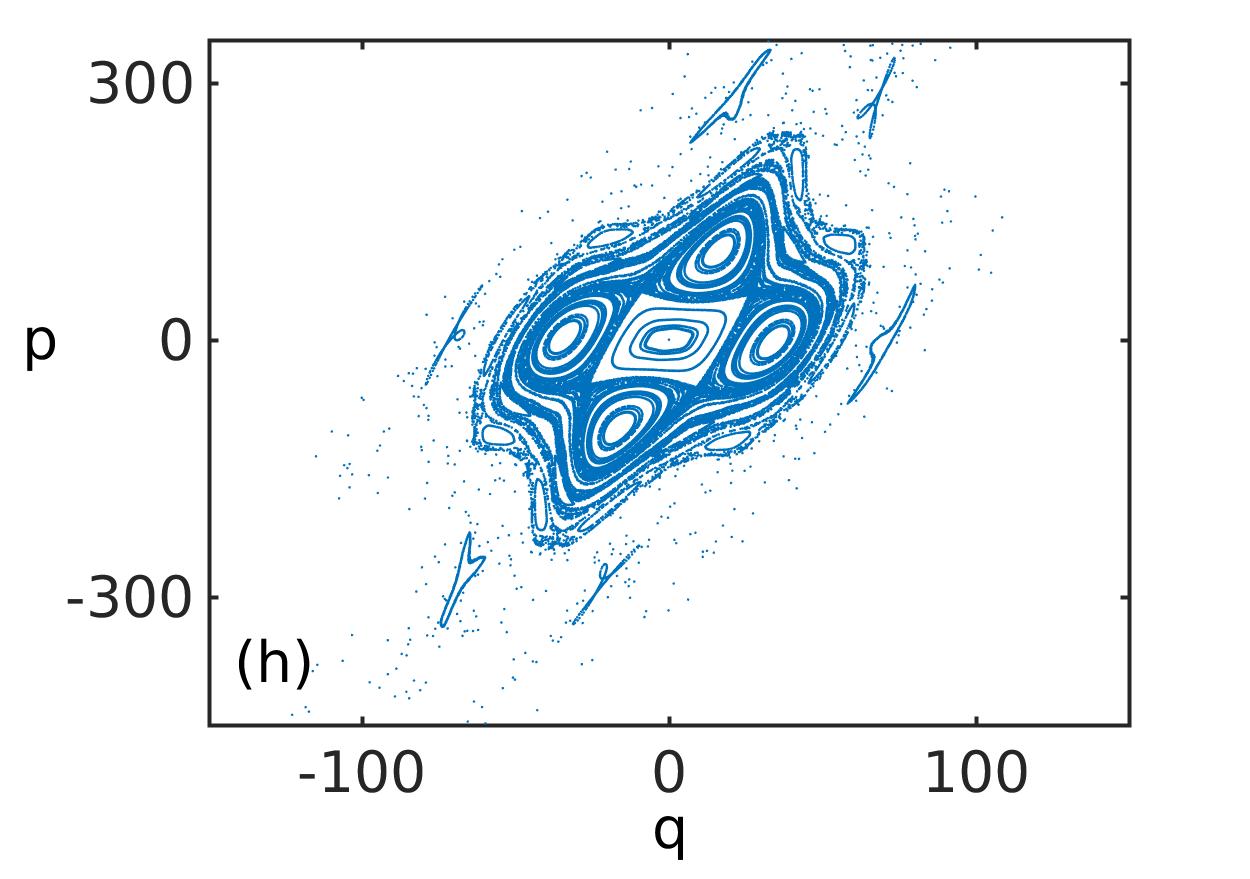}
\includegraphics[width=5cm,height=5cm]{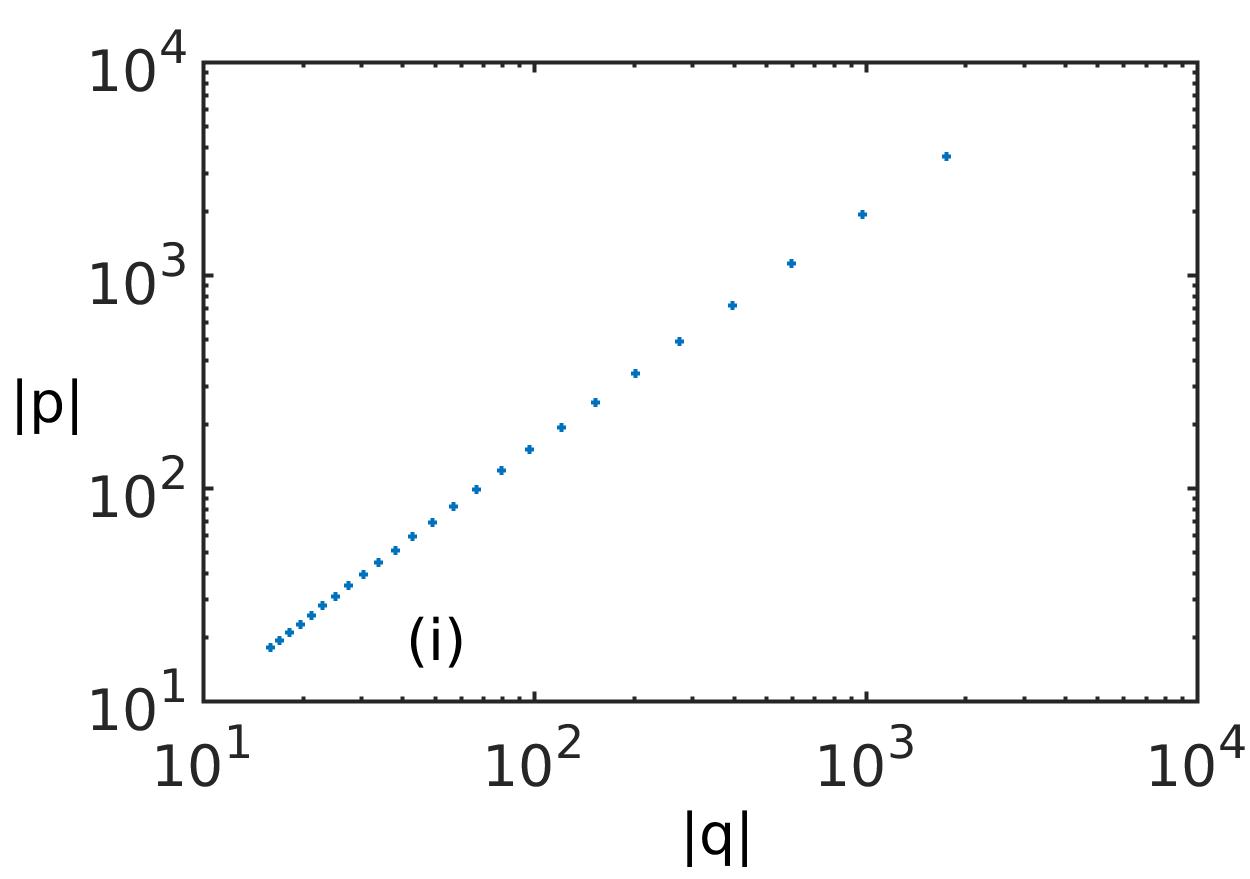}
\caption{\label{fig:fig3} Poincare surfaces of section of phase space $(q,p)$. (a,b) Low frequency $\omega =0.1$
and two different amplitudes $\beta_1=0.1$ and $\beta_1=0.5$, respectively. (c) A somewhat
larger frequency and intermediate amplitude $\omega = 0.3, \beta_1=0.25$.
(d,e,f) show the cases $\omega = 1$ for $\beta_1=0.15,0.2,0.9$ respectively.
(g,h) provide the case of a larger frequency $\omega = 10$ with increasing amplitude $\beta_1=0.1$ and
$\beta_1=0.3$, respectively. (i) shows the magnitude of the coordinates and momenta of
the positions of the dominant period two fixed point of the PSS with varying
$\beta_1$ on a double logarithmic scale. From large to small values in phase space the positions of
the fixed points are indicated by crosses for which $\beta_1$ varies from $0.10$
($|q|=1754,|p|=3593$) to $0.34$ ($|q|=16.0,|p|=17.8$) in steps of $0.01$ for $\omega=1$.
Note the different scales of the subfigures.}
\end{figure}

Figure \ref{fig:fig3}(a) shows the PSS for a comparatively low frequency $\omega=0.1$ and small amplitude 
$\beta_1 = 0.1$ corresponding to the trajectories shown in Figures \ref{fig:fig2}(a,b). All of phase space
appears to be regular i.e. periodic and quasiperiodic motion. Exclusively bounded trajectories are encountered.
The corresponding amplitude of the bounded energy variations during the driving
cycles increases with increasing distance of the trajectories from the stable period one fixed point at the origin (see
Figures \ref{fig:fig2}(a,b)). Note that these statements hold significantly beyond the spatial scales 
shown in Figure \ref{fig:fig3}(a).  
This picture changes if we increase the amplitude to $\beta_1 = 0.5$, see Figure \ref{fig:fig3}(b).
Now a regular island of the phase space centered around the fixed point at the origin is surrounded by
an irregular point pattern. This pattern corresponds to the trajectories analyzed in Figures \ref{fig:fig2} (d-f)
showing a time evolution characterized by an average exponential increase in coordinate and energy space.
Phase space can therefore be divided into two portions: one with bounded and one with unbounded motion.
This picture persists with increasing frequency $\omega$. For a given frequency and small driving amplitude $\beta_1$ 
low energy phase space is regular. With increasing $\beta_1$ the phase space volume of the bounded regular motion
decreases (see below) and gradually the phase space component consisting of 'exponential motion' takes over. 
For a large amplitude $\beta_1=1$ all of phase space consists of exponentially diverging trajectories
and no regular (or chaotic, see below) structures survive. Note that $\beta_1=1$ is the 
case where the lower turning point of the oscillating power of the TPO corresponds to power zero and is therefore
subject to free i.e. unconfined dynamics.

Beyond the above general statements, lets increase the frequency $\omega$ stepwise and see how the
structures in the PSS change. Figure \ref{fig:fig3}(c) shows the PSS for $\beta_1=0.5$ and 
$\omega=0.3$. It exhibits a large-sized centered regular islands around the stable period one fixed at the
origin. Decentered regular islands occur which correspond to period two stable fixed points at $(|q|=33,|p|=33.5)$.
A hierarchical structure of smaller islands around these main ones can equally be observed.
The sequence of PSS in Figures \ref{fig:fig3}(d,e,f) is for an intermediate frequency
$\omega=1$ with increasing amplitude $\beta_1=0.15, 0.2, 0.9$ of the TPO.
Figure \ref{fig:fig3}(e) shows the PSS for $\omega = 1$ and $\beta_1=0.2$.
It exhibits two large separated regular islands around period two fixed points which are off-center. 
The corresponding centered regular island (see inset for its magnification)
with the period one fixed point at the origin is of much smaller size.
The two fixed points of the large decentered regular islands in Figure \ref{fig:fig3}(e) 
describe a two-mode left-right asymmetric motion in the oscillator potential accompanied by
alternating phases of small and large amplitude motion.
Off those large regular islands (see Figure \ref{fig:fig3}(e)) there is an area with many significantly
smaller regular islands interdispersed in a sea of irregular motion. 

We mention that trajectories starting from this mixed portion of phase space, which finally become unbounded
with an exponential increase of their coordinate and momenta (energy), can beforehand show a long period
of stickiness to the neighborhood of small regular islands in this regime. All outer parts of phase space
(see Figure \ref{fig:fig3}(e) with a strongly depleted spreaded point pattern), as discussed-above, 
correspond to exponentially diverging trajectories. Decreasing the amplitude $\beta_1$
leads to an increase of the size of the off-centered large islands. They will then take over a large
part of the completely regular 'low-energy' phase space of the TPO, see Figure \ref{fig:fig3}(d).
In this Figure \ref{fig:fig3}(d) also regular islands around two period three fixed points can be observed.
Increasing the amplitude (see Figure \ref{fig:fig3}(f) for $\beta_1=0.9$ and $\omega =1$)
now shows a completely confined chaotic sea and only tiny regular structures within it.
We note that for $\beta_1 > 0.5$ the above-discussed cusp-like behavior (see corresponding
discussion in the context of Figure \ref{fig:fig1}(a)) of the instantaneous potential of the
TPO for $\beta < 0.5$ is the origin of an unstable behavior and resultingly chaotic sea
in the immediate vicinity of the origin in the corresponding PSS.
Again, outside this confined chaotic sea unbounded exponentially diverging trajectories take over.

Figure \ref{fig:fig3}(g) addresses the case of a much larger frequency $\omega =10$ for small amplitude
$\beta_1 = 0.1$. It shows again an exlusively regular phase space with one large centered main island with the
period one fixed point.
For the same frequency but $\beta_1=0.3$ Figure \ref{fig:fig3}(h) presents a two-component phase space again.
The inner regular island around the period one fixed point shows subislands around a period four
fixed point. The outer part of phase space corresponds exclusively to the exponentially
diverging motion. We conclude that an amazingly broad frequency and amplitude regime is encountered within which
the underlying motion of the TPO is of composite character showing both a regular or mixed regular-chaotic bounded motion and
an exponential unbounded dynamics. Finally Figure \ref{fig:fig3}(i) shows the positions (crosses) 
of the dominant period two stable fixed points of the PSS (period $2T=\frac{4 \pi}{\omega}$ for the TPO)
in phase space with varying amplitude $\beta_1$ and for the frequency $\omega=1$. For $\beta_1=0.10$ 
the stable fixed point occurs at large distances ($|q|=1754,|p|=3593$) and provides the center of an equally
large-sized regular island. With increasing $\beta_1$ it moves inwards i.e. towards the origin
(shown in Figure \ref{fig:fig3}(i) in steps of $0.01$ for $\beta_1$) and is accompanied by a shrinking of the size of
the corresponding regular islands (see also Figures \ref{fig:fig3} (d,e) for $\omega=1$ and
$\beta_1=0.15,0.2$ respectively). Finally at approximately $\beta_1=0.34$ ($|q|=16.0,|p|=17.8$) this regular
island and the stable fixed point disappear. 

\subsection{Quantification of phase space volumes and mean energy gain} \label{sec:vps}

Let us quantify the above observations. To this end we derive computational estimates of the volume $V_{ps}$ of the
phase space that leads to bounded motion. The latter can be purely regular or of mixed regular-chaotic origin
and changes with varying parameters such as the frequency $\omega$ and the amplitude $\beta_1$.
Knowing $V_{ps}(\omega,\beta_1)$ tells us how generic the exponentially diverging dynamics is which occupies
the complementary part of phase space. $V_{ps}(\omega,\beta_1)$ therefore corresponds to the finite and bounded
low energy portion of phase space that 'survives and resists' to the driving in the sense that the energy fluctuations
in the course of the dynamics remain bounded. The unbounded complementary part of phase space is exclusively 
occupied by the exponentially diverging motion and infinite energy growth.

\begin{figure}
\includegraphics[width=9cm,height=7cm]{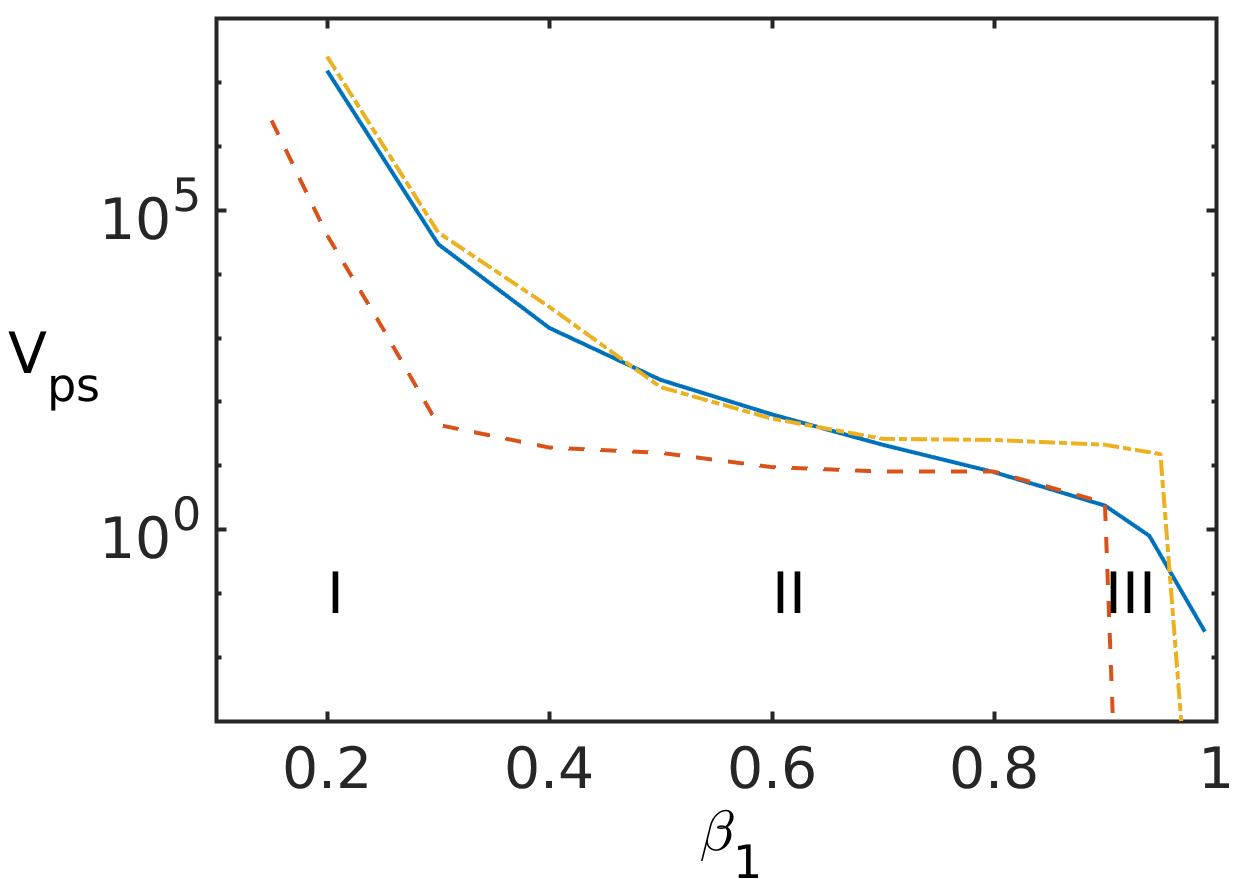} 
\caption{\label{fig:fig4} Volume $V_{ps}$ of phase space in the PSS which corresponds to the
bounded motion with varying oscillation amplitude $\beta_1$ of the TPO.
Full, dashed, and dash-dotted curves correspond to the frequencies $\omega = 0.1,1,10$, respectively.
Note the logarithmic scale for $V_{ps}$.}
\end{figure}

Figure \ref{fig:fig4} shows on a logarithmic scale the volume $V_{ps}$ of phase space in the
stroboscopic PSS with varying oscillation amplitude $\beta_1$ of the TPO.
Three cases of different frequencies $\omega= 0.1, 1, 10$ are presented addressing comparatively low, intermediate and higher
frequencies. $V_{ps}$ has been determined numerically by coarse graining the phase space in the PSS
and approximately measuring the area of the corresponding (regular and chaotic) islands. Since $V_{ps}$
varies over many orders of magnitude (see Figure \ref{fig:fig4}) with varying value of $\beta_1$ we are
predominantly interested in a crude estimate of its value.
Three different regimes I,II and III can be distinguished in Figure \ref{fig:fig4}.
For small amplitudes $\beta_1 \lesssim 0.25$ (regime I) the volume $V_{ps}$ is extremely sensitive to the 
value of the amplitude $\beta_1$ of the TPO. It covers many orders of magnitude e.g. decreasing from
$2.5 \times 10^{6}$ to approximately
$40$ when $\beta_1$ increases only from $0.15$ to $0.3$ for the case of $\omega =10$. This clearly demonstrates
that for small amplitudes $\beta_1$ the phase space is rapidly taken over by regular and bounded motion in agreement with the
above-discussed PSS (see Figures \ref{fig:fig3}(a,d,g)). The dependence of $V_{ps}$ on $\beta_1$ becomes
much weaker in the intermediate regime $0.3 \lesssim \beta_1 \lesssim 0.9$ (regime II),
where e.g. for the case $\omega=1$ almost a plateau-like structure can be observed.
While the volume $V_{ps}$ of bounded motion does not change as much as in regime II as it does in regime I,
the character of the confined motion could severly change. This is demonstrated in Figures \ref{fig:fig3}(f,h).
For the low frequency $\omega =0.1$ a purely regular island of bounded motion occurs whereas for an
intermediate frequency $\omega =1$ a bounded chaotic sea is encountered. Finally, in regime III for
$1 \ge \beta_1 \gtrsim 0.9$ we observe again a rapid decrease of $V_{ps}$ for $\beta_1$ increasing from $0.9$ to $1$.
In the limit $\beta_1=1$ we arrive at $V_{ps} =0$ and all of phase space consists of exponentially
diverging trajectories. Note that the corresponding transition needs to be by no way smooth. Its details
depend on the destruction of the last invariant spanning manifolds which could, as it is for example the case
for $\omega=1$ and $\beta_1 > 0.9$ (see confined chaotic sea in Figure 
\ref{fig:fig3}(f)), lead to a sudden transition from a finite bounded volume to zero volume.

\begin{figure}
\includegraphics[width=9cm,height=7cm]{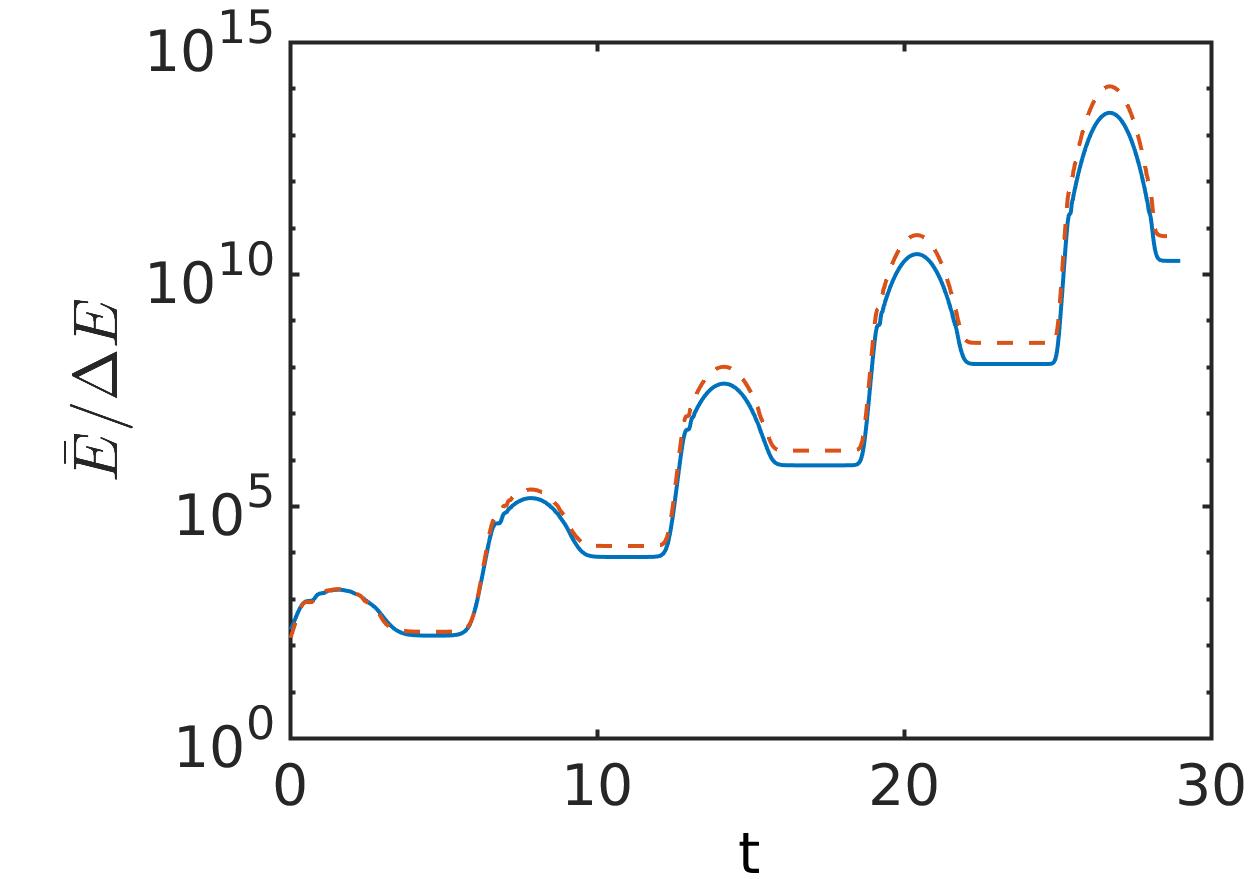} 
\caption{\label{fig:fig5} Mean energy $\bar{E}$ (full line) and its standard deviation $\Delta E$ (dashed line) as a function
of time for a statistical ensemble of trajectories sampled from the exponentially accelerating part of the phase space for $\omega = 1$
and $\beta_1 = 0.9$.}
\end{figure}

It is instructive to consider not only individual trajectories in the regime of exponential growth 
but the behavior of the mean of the energy $\bar{E}$ and the corresponding standard deviation
$\Delta E$ for an ensemble of trajectories
with randomly chosen initial conditions in this regime. We note, that these trajectories are sensitive
to the initial conditions as it holds for any chaotic trajectory. Figure \ref{fig:fig5} shows these two 
quantities $\bar{E},\Delta E$ as a function of time for 100 trajectories of the phase space for $\omega =1, \beta_1=0.9$
(see also Figure \ref{fig:fig3}(f)). The envelope behavior shows an exponential increase of both $\bar{E},\Delta E$
where the standard deviation is slightly larger than the mean. The cycles of the 'pumping' of the TPO (the period of the
oscillator is here $2 \pi$) are clearly visible. Note the extraordinary large energy gain per cycle which amounts
to two orders of magnitude in the energy per cycle. As such this oscillator is therefore an extremely efficient 
'accelerator' or energy pumping 'device'. We remark that our numerical simulations have been performed for larger
times than the ones shown in Figure \ref{fig:fig5} and support this picture. However, with increasing propagation
time the computational time to integrate one further oscillator period grows substantially and therefore limits
the time-evolution to be probed.

As already indicated in subsection \ref{sec:trajectories} in the
context of the discussion of individual trajectories we observe a rapid energy change
for the first half-cycle $nT < t < (n+1/2)T$ ($n$ is a positive integer) for which the strong confinement of
the TPO obeys $\beta (t) > \beta_0$. For a substantial part of the second
half-cycle $(n+1/2)T < t < (n+1)T$ with the weaker confinement $\beta (t) < \beta_0$ the energy 
remains approximately unchanged (on the logarithmic scale of Figure \ref{fig:fig5}).
This picture provides however an incomplete description of the growth of the energy.
Indeed, the dynamics and time-evolution of the energy is asymmetric with respect to the 
turning-points of the TPO $(n+1/4)T, (n+3/4)T$
for both the first and second half-cycle. This asymmetry is particularly pronounced for the second
half cycle. Specifically, during the first quarter cycle $nT < t < (n+1/4)T$ the energy increases
whereas it decreases to a slightly lesser extent during the second quarter cycle $(n+1/4)T < t < (n+1/2)T$.
At the beginning of the second half cycle the energy saturates and remains approximately constant, but does so not
for all of the second half cycle. Instead it raises substantially again towards the end of the
second half cycle, i.e. within the last quarter cycle. This strong asymmetry leads to the net energy gain of the TPO.

\subsection{High frequency regime} \label{sec:hfr}

Finally let us discuss the dynamics of our TPO in the high frequency limit $\omega \rightarrow \infty$.
Obviously in this limit the largest frequency in the system is the one associated with the time-changing power of the TPO.
A way to describe this situation is to average the rapidly oscillating potential over a
period of the oscillation thereby obtaining an effective time-independent potential \cite{Moiseyev,Diakonos,
Gaerttner,Koch,Fishman1,Fishman2}

\begin{equation}
{\cal{V}}_{eff} = \frac{1}{T} \int_{0}^{T} \alpha q^{2 \beta(t)} dt \label{Veff_1}
\end{equation}

where we remind the reader that $\beta(t)=\beta_0+\beta_1 sin(\omega t)$. Due to the integration in
eq.(\ref{Veff_1}) ${\cal{V}}_{eff}$ is independent of the frequency $\omega$ and depends solely on $\beta_0,\beta_1$.
The question is how this autonomous (energy-conserving) one-dimensional potential, which leads to exclusively
integrable i.e. periodic motion, looks like. Or, in other words, which features of the different instantaneous power-law potentials
reflect themselves in the time-averaged high frequency limit leading to ${\cal{V}}_{eff}$.
For potentials of a product form concerning
their time and spatial dependence such as the parametrically driven
harmonic oscillator $V_{po}(t) = \beta (t) q^2$ the shape of the time-averaged potential is the same as
the original one. This is certainly not the case for our TPO.
Rescaling time under the integral and accounting for the regularization of the power-law
potential eq.(\ref{Veff_1}) becomes

\begin{equation}
{\cal{V}}_{eff} = \frac{\alpha}{2 \pi} \left( q^2 + C \right)^{\beta_0} \int_{0}^{2 \pi} \left(q^2 + C \right)^{\beta_1 sin(t)} dt
\label{Veff_2}
\end{equation}

Exploiting the fact that the subintegrals in the intervals $[0,\frac{\pi}{2}]$ and $[\pi,\frac{3\pi}{2}]$ equal 
those in the intervals $[\frac{\pi}{2},\pi]$ and $[\frac{3\pi}{2}, 2\pi]$, respectively, as well as employing the
substitution $u=\beta_1 sin(t)$ results in the following representation of the effective potential

\begin{equation}
{\cal{V}}_{eff} = \frac{\alpha}{\pi \beta_1} \left( q^2 + C \right)^{\beta_0} \int_{0}^{\beta_1}
\left(\left(q^2 + C \right)^{u} - \left(q^2 + C \right)^{-u} \right) \frac{1}{\sqrt{1 - \left(\frac{u}{\beta_1}\right)^2}} du
\label{Veff_3}
\end{equation}

The peculiarity of both representations of ${\cal{V}}_{eff}$ in eqs.(\ref{Veff_2},\ref{Veff_3}) is the fact that
the integration is performed with respect to the exponent of the integrand. As a result the integral
cannot be expressed in terms of simple known analytical functions i.e. it has to be performed numerically.
This goes
hand in hand with our previous remark that in general any analytical approach to the TPO seems to be very
difficult due to its time-dependent, in general fractional, exponent. 


\begin{figure}[H]
\includegraphics[width=7cm,height=6cm]{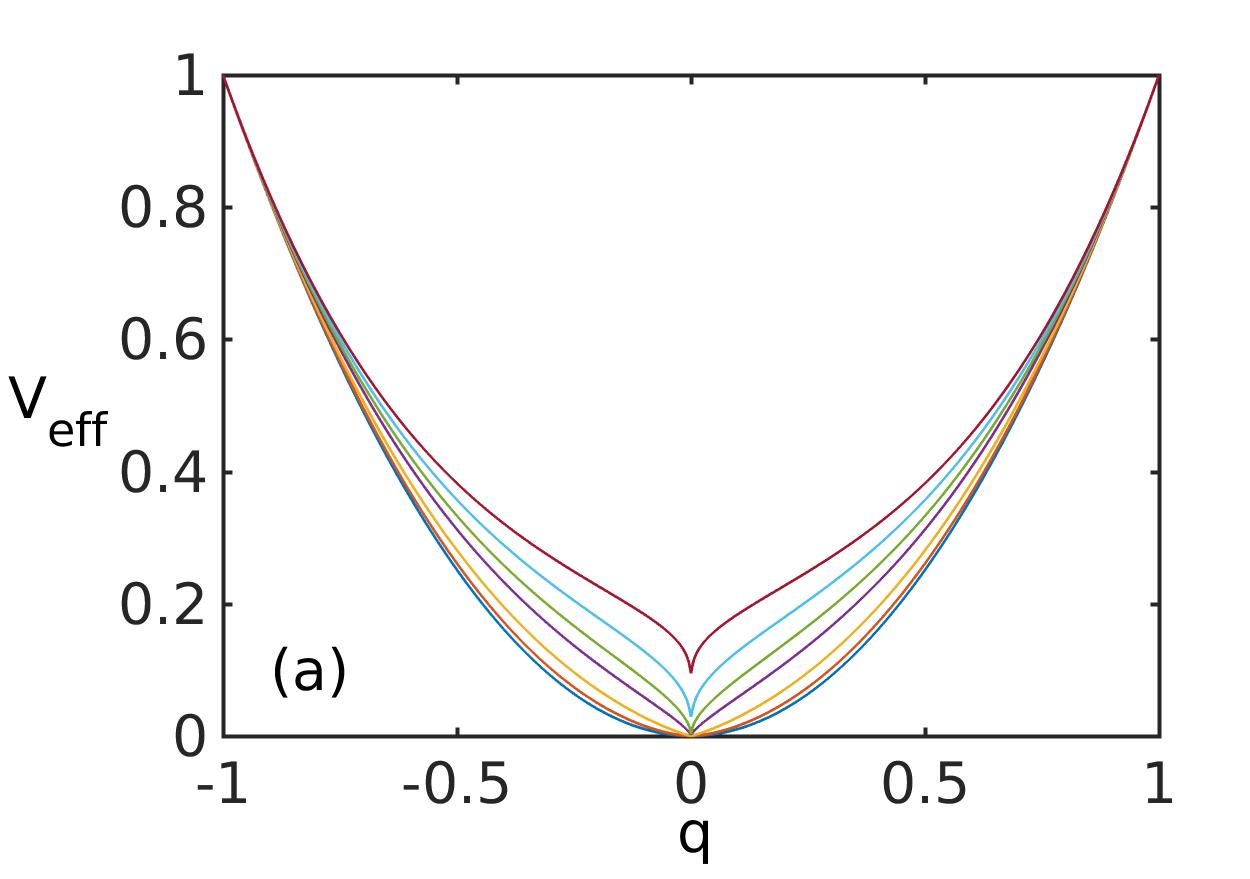} 
\includegraphics[width=7cm,height=5.75cm]{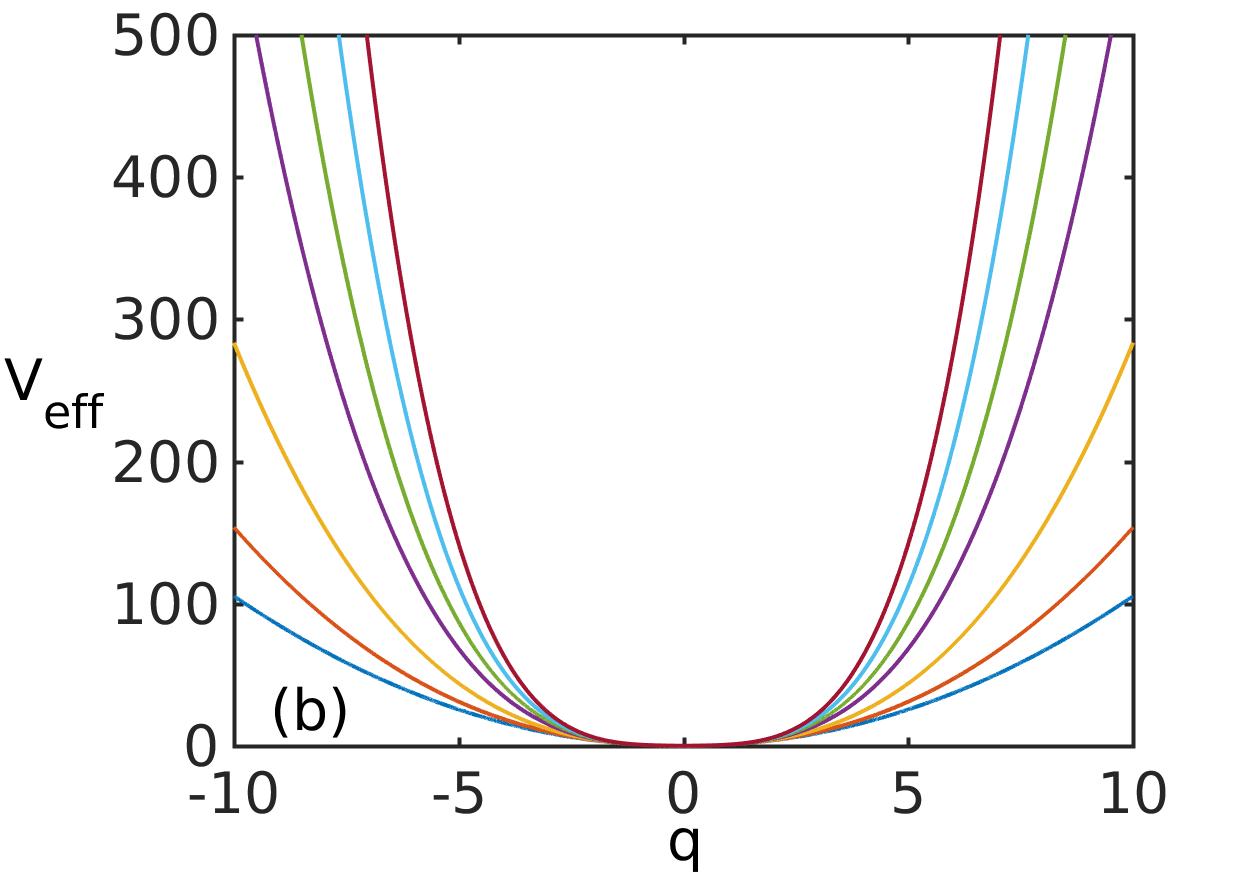} 
\caption{\label{fig:fig6} Effective potential ${\cal{V}}_{eff}$ in the high frequency limit (see text) 
(a) in the vicinity of the origin and (b) for a larger $q$ interval. The assignment of the values 
$\beta_1 = 0.1, 0.3, 0.5, 0.7, 0.8, 0.9, 0.99$ corresponds to the curves from bottom to top.}
\end{figure}

\begin{figure}[H]
\begin{center}
\includegraphics[width=7cm,height=7cm]{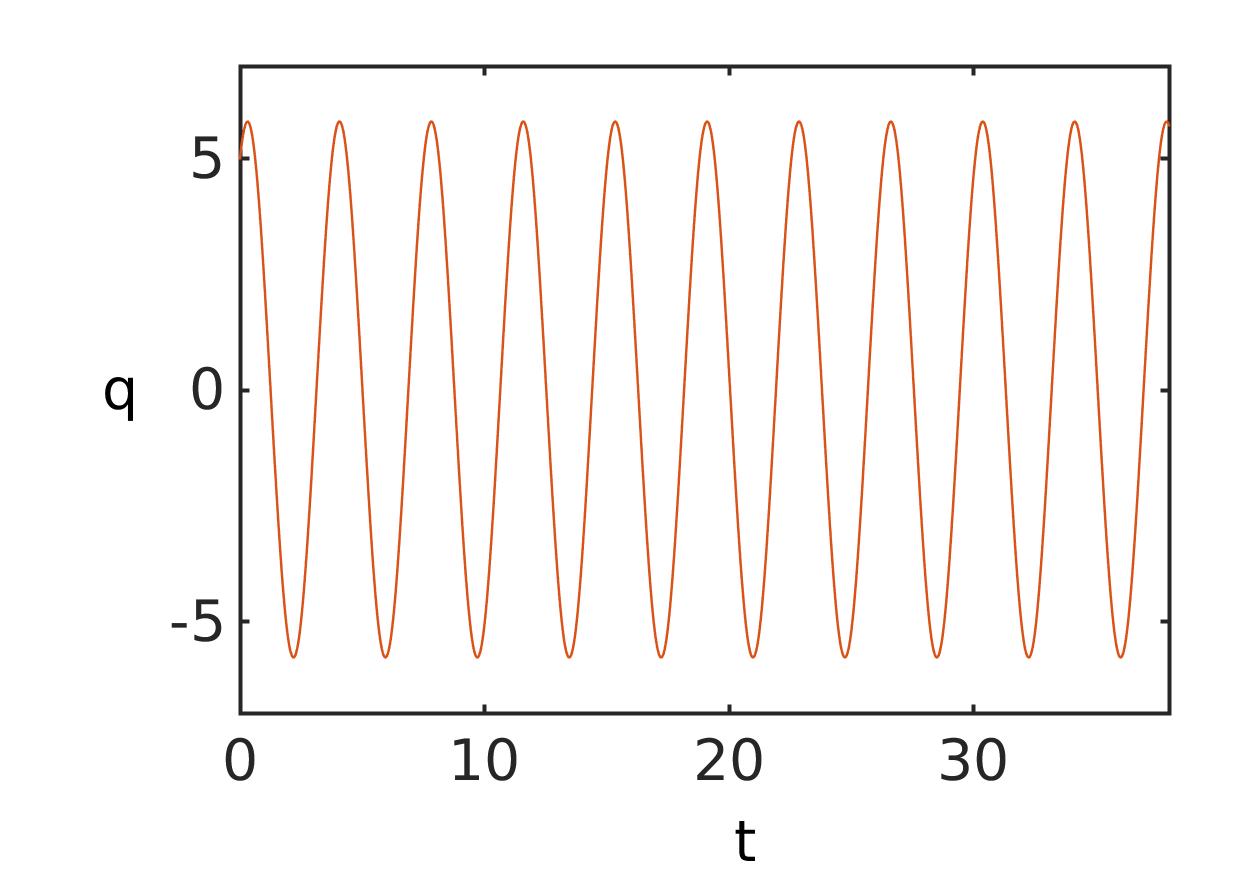} 
\end{center}
\caption{\label{fig:fig7} Trajectories $q(t)$ in coordinate space with initial condition
$q(t=0)=5,p(t=0)=5$ for both the TPO and the time-independent effective potential (see eqs.(\ref{Veff_2},
\ref{Veff_3})). The two curves are indistinguishable i.e. on top of each other on the resolution of the figure.
The TPO parameters chosen are $\omega=1000, \beta_1=0.3$.}
\end{figure}

Figures \ref{fig:fig6}(a,b) show the effective  potential ${\cal{V}}_{eff}$ in the high frequency limit,
as obtained by a numerical integration.
Let us first focus on the behavior close to the origin, which is shown in subfigure \ref{fig:fig6} (a). 
With increasing $\beta_1$ ranging from $0.1$ to $0.7$ the flat bottom of the potential around the origin
develops into a linear rise. Further increasing the distance from the origin this linear behavior
turns into a quadratic and subsequently higher order one. This happens because the linear scaling dominates in the
integral for the effective potential ${\cal{V}}_{eff}$ for very short distances as compared to the quadratic
one. It is however only 'accessible' within the TPO if the amplitude $\beta_1$ is sufficiently large.
For $\beta_1 > 0.7$ the behavior close to the origin is characterized by a cusp-like dip, which can be 
assigned to the dominance of the powers of the TPO being sublinear (see also the corresponding snapshots of the 
static potential of the TPO in Figure \ref{fig:fig1}(a)) and possesses for $C=0$ diverging derivatives at
the origin. For larger distances from the origin the potential changes its second derivative from
negative to positive. All potentials become degenerate at $q = \pm 1$ independently of the value of $\beta_1$.
Figure \ref{fig:fig6}(b) shows ${\cal{V}}_{eff}$ on a larger spatial scale where a clear ordering
with respect to the strength of the asymptotic ($q \rightarrow \infty$) confinement is visible. Larger
amplitudes $\beta_1$ correspond to a stronger asymptotic confinement in this high frequency limit.
Finally let us compare the TPO dynamics and the dynamics in the time-independent effective potential
${\cal{V}}_{eff}$ (see eqs.(\ref{Veff_2},\ref{Veff_3})). Figure \ref{fig:fig7} shows typical trajectories
for the initial condition ($q(t=0)=5,p(t=0)=5$) for both cases, where $\omega=1000$ has been chosen for the
TPO. The differences w.r.t. their dynamics not being visible for the finite resolution of Figure \ref{fig:fig7} amount
to maximally 0.1 percent for the propagation time shown. This confirms the validity of the high frequency
picture derived above. Let us shed some light on the behaviour of the TPO with increasing frequency towards the high 
frequency limit. The period $T$ of the oscillations for our specific trajectory
is $T=3.7878$ for $\omega=10$ and becomes $T=3.7590$ for $\omega=1000$ whereas the effective potential
provides the value $T=3.7588$. We note that the corresponding time-independent harmonic oscillator
($\alpha=1, \beta_0=1, \beta_1=0$) amounts to $T=4.4429$. These comparatively small deviations in the frequency
should however not obscure the fact that there are major deviations between the cases $\omega=10$ for the
TPO and the effective potential dynamics for the trajectory $q(t)$ which amount up to 12 percent.

\section{Conclusions and outlook} \label{sec:conclusions}

We have explored the nonlinear dynamics of the driven power-law oscillator with a focus on
the analysis of the phase space structures and the possibility of energy gain of the oscillator.
When comparing to the existing models for oscillators in the literature, the TPO appears to be of
unusual appearance. Opposite to the kicked or parametrically driven oscillators which keep their
shape during driving, the TPO continuously changes its shape in the course of a period of the periodic
driving. Choosing as an equilibrium the harmonic oscillator, during one half-period of the driving
the confinement strength of the anharmonic TPO is larger and during the other half-period the
power of the confining potential is smaller when compared to the harmonic oscillator.
With increasing amplitude of the oscillating power the oscillator potential can become linear
or even inverse its curvature, such that the confining properties become correspondingly weak.
In the latter case a cusp appears in the instantaneous potential that
leads to kicks in the dynamics particularly relevant for low velocity trajectories near the origin.

Our computational study has shown that for sufficiently large amplitude the TPO exhibits exponential
acceleration and energy gain within the regime of phase space and time-scales considered here.
We conjecture that this holds even for an infinitely large part of the unbounded phase space. The structure
of the finite low-energy part of the phase space around the origin
exhibits bounded motion which can be completely regular (small amplitude, arbitrary frequency) or
mixed regular-chaotic (intermediate amplitude) up to the case of a dominantly chaotic phase space
(intermediate amplitude and frequency). We unraveled an intriguing mechanism of energy gain on the
level of individual trajetories which links to the
power-law pumping during the cycling of the TPO. Performing a corresponding statistical analysis
we could assign the different periods of energy flow in and out of the oscillator to the phases of
the external driving with the net result of an exponential energy gain. The exponential acceleration
takes place in an amazingly broad parameter regime w.r.t. the frequency and amplitude of the TPO.
The volume of phase space which corresponds to the bounded motion is extremely sensitive to the amplitude
of the driving varying over many orders of magnitude and showing a rich structure. 
We remark that the observed exponential acceleration could potentially also be related to heating which has very
recently been found to occur in periodically driven many-particle systems \cite{Bukov} showing
different 'temperatures' after thermalization for slow and fast driving.
Finally, in the high frequency limit, an effective time-averaged and static potential has been determined
which combines interesting features, such as the near origin cusp (for sufficiently large driving
amplitude) and a very strong effective confinement at large distances. The exact dynamics 
reflects this high frequency behavior only for large values of the frequency.

Although our driven power law oscillator is a novel kind of oscillator with
peculiar properties and as such, to our opinion, of interest on its own, let us briefly
address possible experimental realizations. Of course, no experimental setup will ever
lead to an infinitely spatially extended confinement such that any concrete realization will
always probe the TPOs dynamics on finite time scales. This holds
in particular w.r.t. the component of the exponentially diverging
motion. Since the trapping technology of cold ions and especially of neutral atoms has advanced
enormously during the past decades \cite{Schneider,Hughes,Wilpers,Pethick} there exists a huge flexibility
concerning the design and the time-dependent change also of the shape of these traps.
One can therefore use a gas of e.g. thermal atoms for which interaction effects are negligible and 
'pump' the oscillator according to the TPO in order to finally perform absorption spectroscopy
at different time instants and resultingly observe the expansion dynamics of the atomic cloud.
This is a microscopic example, and does in particular not exclude the possibility of probing
the TPO at hand of a macrosopic i.e. mechanical setup. As can be seen from the time-evolution
of the energy of the oscillator our TPO is an extremely efficient
'accelerator' and one could in principle imagine its application as a few-cyle trap accelerator
efficiently switching between different energy regimes.

As an outlook to the future, it might be interesting but very challenging to provide an analytical approach
to the TPO in some of the accessible parameter regimes. However, it is not clear whether these plans are
destined to fail, since even the simplest integrals are not expressable in terms of standard functions because it is
typically the time-dependent exponent which is to be integrated. Still, a further more detailed analysis
and understanding of the rich structure of the TPO is certainly desirable and, if not analytically, then
computationally an interesting endavor. Going one step further and analyzing the quantum TPO via e.g. Floquet
theory and identifying the characteristics of its Floquet spectrum is an intriguing perspective.

\section{Acknowledgments}

This work has been in part performed during a visit to the Institute for Theoretical Atomic, Molecular and Optical Physics (ITAMP)
at the Harvard Smithsonian Center for Astrophysics in Cambridge, Boston, whose hospitality is gratefully acknowledged.
The author thanks B. Liebchen for a careful reading of the manuscript and valuable comments.

\end{document}